\documentclass[10pt,sigconf,letterpaper]{acmart}

\usepackage{booktabs} %
\usepackage{graphicx}   %
\usepackage{mathtools}
\usepackage{amsmath}
\usepackage{color, colortbl}
\usepackage[nolist]{acronym}
\usepackage{multirow}
\usepackage{euscript}
\usepackage{adjustbox}
\usepackage{array}
\usepackage{appendix}
\usepackage[tableposition=bottom]{caption}
\usepackage{subfig}
\usepackage{placeins}
\usepackage[noabbrev]{cleveref}

\usepackage{hyperref}
\def\Snospace~{\S{}}

\hypersetup{
    colorlinks,%
    citecolor=[rgb]{0.7,0,0},
    filecolor=[rgb]{0.2,0.2,0.2},
    linkcolor=[rgb]{0,0,0.7},
    urlcolor=[rgb]{0.7,0,0},
}

\definecolor{Gray}{gray}{0.9}
\newcommand{\Vs}[1]{\text{\textit{#1}}}  %

\newcolumntype{C}{>{\centering\arraybackslash}m{4.5em}}
\newcolumntype{D}{>{\centering\arraybackslash}m{3.5em}}
\newcolumntype{M}[1]{>{\centering\arraybackslash}m{#1}}
\newcolumntype{P}[1]{>{\centering\arraybackslash}p{#1}}
\newcolumntype{N}{@{}m{0pt}@{}}

\setcopyright{none}

\renewcommand\footnotetextcopyrightpermission[1]{} %

  \acmDOI{}
  \acmISBN{}
  \acmConference{Technical Report}{Dec 2024}{Marina del Rey, CA, USA}
  \acmYear{}
  \copyrightyear{2024}
  \acmArticle{}
  \acmPrice{}

\makeatletter
  \newcommand\EatSpacesHack{\@bsphack\@esphack}
\makeatother

  \newcommand\oldreviewfix[1]{\EatSpacesHack}

  \makeatletter
    \renewcommand\section{\@startsection{section}{1}{\z@}%
      {-.25\baselineskip \@plus -2\p@ \@minus -.2\p@}%
      {.1\baselineskip}%
      {\ACM@NRadjust\@secfont}}
    \renewcommand\subsection{\@startsection{subsection}{2}{\z@}%
      {-.15\baselineskip \@plus -2\p@ \@minus -.2\p@}%
      {.05\baselineskip}%
      {\ACM@NRadjust\@subsecfont}}
    \renewcommand\subsubsection{\@startsection{subsubsection}{3}{\z@}%
      {-.15\baselineskip \@plus -2\p@ \@minus -.2\p@}%
      {-2\p@}%
      {\ACM@NRadjust{\@subsubsecfont\@adddotafter}}}
   \makeatother
  \newcommand{\RelaxFloats}{
  	\renewcommand{\topfraction}{0.9}
  	\renewcommand{\floatpagefraction}{0.9}
  	\renewcommand{\textfraction}{0.1}
  }
  \RelaxFloats

\acrodefplural{AS}[ASes]{Autonomous Systems}
\acrodefplural{VP}[VPs]{Vantage Points}
\acrodef{NAT}[NAT]{Network Address Translation}
\acrodef{CG-NAT}[CG-NAT]{Carrier-Grade NAT}
\acrodefplural{CDN}[CDNs]{Content Delivery Networks}
\acrodef{RIR}[RIR]{Regional Internet Registry}
\acrodefplural{RIR}[RIRs]{Regional Internet Registries}

\begin{document}
\title{Measuring Partial Reachability in the Public Internet}

  \setlength{\dblfloatsep}{-5pt}
  \addtolength{\abovecaptionskip}{-4pt}
  \setlength{\dbltextfloatsep}{0pt}
  \setlength{\abovedisplayskip}{1.5pt plus 1pt minus 1pt}
  \setlength{\belowdisplayskip}{1.5pt plus 1pt minus 1pt}

  \settopmatter{authorsperrow=4}
  \author{Guillermo Baltra}
  \affiliation{%
    \department{USC/ISI}
    \city{Marina del Rey}
    \state{California}
    \country{USA}
  }
  \email{baltra@isi.edu}

  \author{Tarang Saluja}
  \affiliation{%
    \department{Swarthmore College}
    \city{Swarthmore}
    \state{Pennsylvania}
    \country{USA}
  }
  \email{tsaluja1@swarthmore.edu}

  \author{Yuri Pradkin}
  \affiliation{%
    \department{USC/ISI}
    \city{Marina del Rey}
    \state{California}
    \country{USA}
  }
  \email{yuri@isi.edu}

  \author{John Heidemann}
  \affiliation{%
    \department{USC/ISI}
    \city{Marina del Rey}
    \state{California}
    \country{USA}
  }
  \email{johnh@isi.edu}

  \renewcommand{\shortauthors}{Baltra et al.}

\begin{abstract}
The Internet provides global connectivity
  by virtue of a public core---the
  routable public IP addresses
  that host services and to which cloud,
  enterprise, and home networks connect.
Today the public core faces many challenges to uniform, global reachability:
  firewalls and access control lists,
  commercial disputes that stretch for days or years,
  and government-mandated sanctions.
We define two algorithms to detect partial connectivity:
  \emph{Taitao} detects
  \emph{peninsulas}  of persistent, partial connectivity,
  and \emph{Chiloe} detects \emph{islands},
  when one or more computers are partitioned
    from the public core.
These new algorithms apply to existing data collected by multiple
  long-lived measurement studies.
\emph{We evaluate these algorithms with rigorous measurements
  from two platforms}:
  Trinocular, where 6 locations observe 5M networks frequently,
  RIPE Atlas, where 10k locations scan the DNS root frequently,
and validate adding a third:
  CAIDA Ark, where 171 locations traceroute to millions of networks daily.
Root causes suggest that
  most peninsula events (45\%) are routing transients,
  but most peninsula-time (90\%) is due to long-lived events (7\%).
We show that the concept of \emph{peninsulas and islands can improve
  existing measurement systems}.
They identify measurement error
  and persistent problems
  in RIPE's DNSmon that are $5\times$ to $9.7\times$ larger
  than the operationally important changes of interest.
They explain previously contradictory results in
  several outage detection systems.
\emph{Peninsulas are at least as common as Internet outages},
  posing new research direction.
\end{abstract}

\begin{acronym}[AS]
\acro{AS}{Autonomous System}
\acro{RDNS}{Reverse DNS}
\acro{VP}{Vantage Point}
\end{acronym}

\maketitle

\vspace*{-1ex}

\section{Introduction}
	\label{sec:introduction}

Internet users would like an Internet that either works or it doesn't.
They understand that the ``Internet doesn't work''
  when they are too far from the wifi access point,
  if their company's router reboots,
  or when they lose power.
However, most of the time
  they expect to be connected to the Internet
  and able to reach any public computer.

Unfortunately, the Internet is not so simple---this paper
  shows that \emph{partial reachability is a fundamental part of the Internet
  and is surprisingly common}.
While top-100 websites are hosted at many points-of-presence
  and are nearly always reachable,
  partial reachability is also common,
  where some destinations on the Internet
   can be reached from some sources, but not from others.
In fact, we will show that in IPv4
  \emph{partial reachability on the Internet is at least as common as
  an outage (complete Internet unreachability)} (\autoref{sec:peninsula_frequency}).

Although partial reachability
  may seem unimportant (who needs to get to \texttt{obscure.example.com}
  when Youtube and Facebook beckon),
  we will show that partial reachability can obscure
  existing measurement systems,
  cluttering their basic results and hiding what is important.
Reexamining widely used RIPE DNSmon in light of partial reachability
  (\autoref{sec:dnsmon}),
   we show that its observations of
   persistent high query loss (5--8\% to the DNS Root~\cite{RootServers16a})
   are mostly measurement error and
   persistent partial connectivity.
These factors are $5\times$ and $9.7\times$ (IPv4 and v6)
  larger than operationally important signals.
Our analysis also helps resolve uncertainty in
  Internet outage detection (\autoref{sec:local_outage_eval}),
  clarifying ``corner cases''
  due to conflicting observations~\cite{Schulman11a,quan2013trinocular,Shah17a,richter2018advancing,guillot2019internet}.

Persistent partial unreachability can also occur because of policy choices.
Firewalls protect one's internal network from the Internet,
  and Access Control Lists sometimes serve as partial firewalls,
  preventing access from parts of the Internet while admitting others.
Although some countries are well known for strict
  policies on international network access~\cite{Anonymous12a}
  particularly during unrest~\cite{Dainotti11a,Gill15a,Dahir18a,Griffiths19a,Jahan24a},
  our examination of global reachability shows that
  95 U.S.-based ASes
  that have long-term blocks on international traffic to 429 /24 blocks (\autoref{sec:detecting_country_peninsulas}).

Partial reachability has been observed before~\cite{Dhamdhere18a},
  and several systems were designed to mitigate partial
  reachability by relaying through a third location~\cite{andersen2001resilient,katz2008studying,katz2012lifeguard}.
While these prior systems reported the existence of partial reachability,
  we are the first to search for it Internet-wide
  and quantify how often and how long partial reachability exists
  in the general IPv4 Internet (\autoref{sec:evaluation}).

\begin{figure*}
  \begin{minipage}[b]{.31\linewidth}
    \mbox{\includegraphics[width=\linewidth,trim=210 300 220 145,clip]{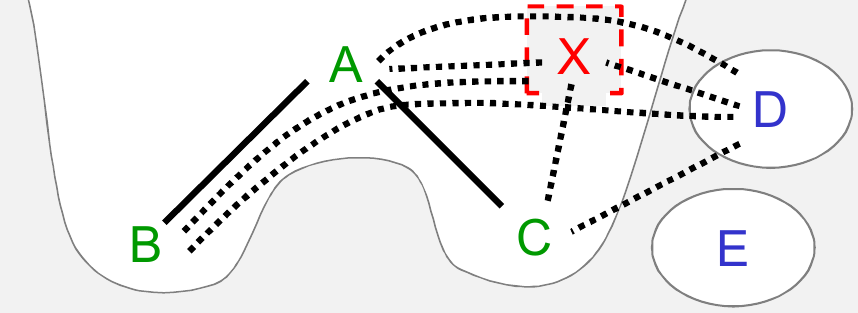}}
    \caption{$A$, $B$ and $C$ are the connected core,
      with $B$ and $C$ peninsulas;
      $D$ and $E$ islands;
      $X$ is out.}
      \label{fig:term_concept}
      \vspace{4ex}
  \end{minipage}
\hspace{1ex}
  \begin{minipage}[b]{.31\linewidth}
    \includegraphics[width=\linewidth]{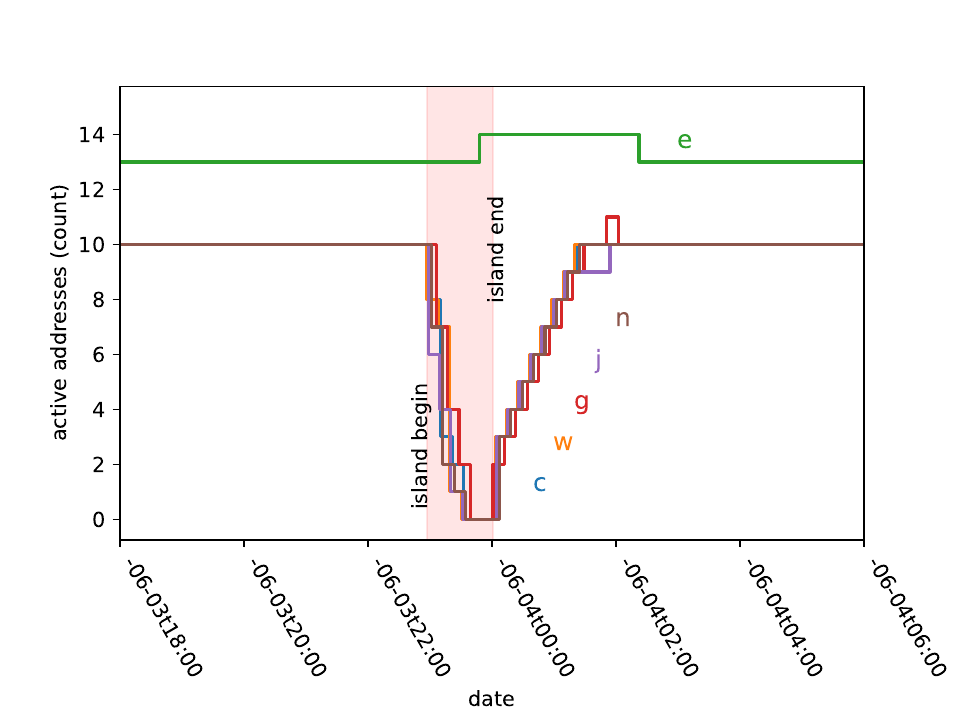}
    \caption{Estimates reachable addresses for an island %
      starting 2017-06-03t23:06Z and lasting 1\,hour.}%
        \label{fig:a28all_417bca00_accum}
  \end{minipage}
\hspace{1ex}
  \begin{minipage}[b]{.31\linewidth}
\includegraphics[width=\linewidth]{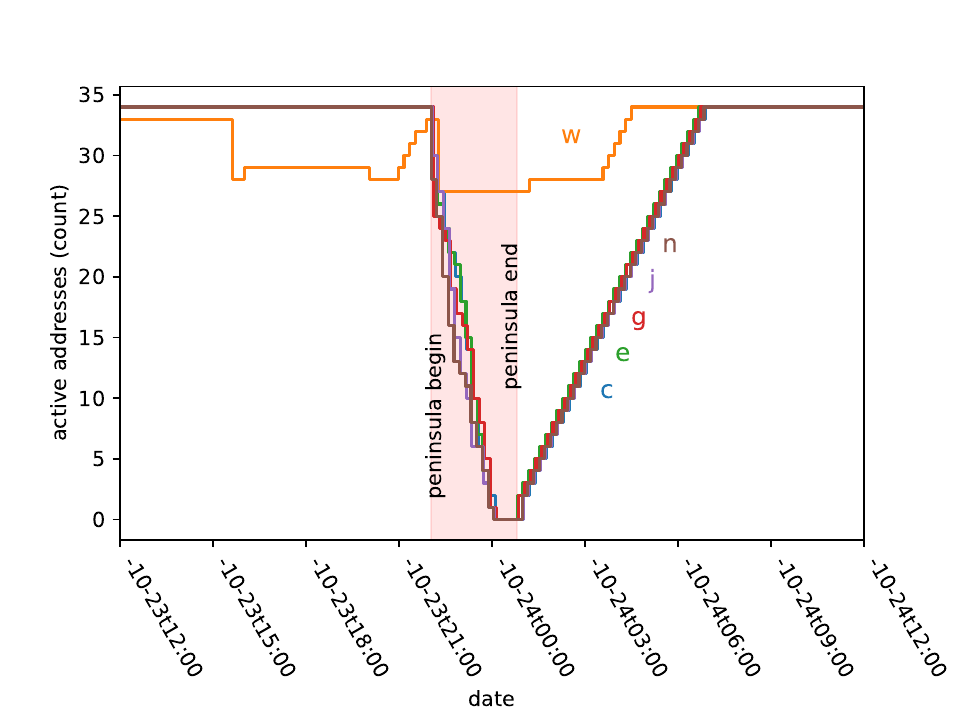}
\caption{Estimates of reachable addresses for a peninsula %
  starting 2017-10-23t22:02Z, lasting  3\,hours.}
    \label{fig:a30all_50f5b000_accum}
  \end{minipage}
\end{figure*}

The first contribution of our paper is to
  define two algorithms that detect partial connectivity (\autoref{sec:design}).
\emph{Taitao} detects peninsulas
  that often result from peering disputes or long-term firewalls.
Our second algorithm, \emph{Chiloe}, detects islands.
We evaluate these algorithms
  using data from two different, long-running measurement systems.
We look for partial outages in Trinocular,
  reexamining 3 quarters of data taken across 4 years
  from 6 \acp{VP}
  to 5M global networks.
We also reexamine 3 years of RIPE Atlas data
  taken from about 10k VPs to 13 different locations.

Our second contribution is to validate these algorithms.
We compare data from the above two systems with each other
  and against a third system, CAIDA Ark,
  where 171 \acp{VP} scan millions of networks, daily~\cite{CAIDA07b}.
While each of these three systems have different
  numbers of VPs and destinations,
  we see very good recall (0.94)
  and reasonable precision (at least 0.42 when strictly
  interpreting different systems, or 0.82 with a broader interpretation)
  in \autoref{sec:taitao_validation}.
We also examine the consistence of our results
  when we reduce the number of observers.
We find that combinations of any three independent \acp{VP}
  provide a result that is statistically indistinguishable from the asymptote
  (\autoref{sec:peninsula_frequency}).
Our algorithms
  provide consistent results
  and offer reproducible and useful estimates
  of Internet reachability and partial connectivity.

Our final contribution
  is to apply these algorithms to the Internet as a whole.
We report on the frequency (\autoref{sec:peninsula_frequency}),
  duration (\autoref{sec:peninsula_duration}),
  and location (\autoref{sec:peninsula_locations})
  of peninsulas.
We show that country-level peninsulas occur (\autoref{sec:country_peninsulas}),
  suggesting peninsulas occur as an organizational policy.
We also report on island frequency (\autoref{sec:how_common_are_islands}),
  duration (\autoref{sec:islands_duration}),
  and sizes (\autoref{sec:islands_sizes}).
Finally,
  we show that the concept of \emph{peninsulas and islands can improve
  existing measurement systems}.
We described early how measurement errors due to
  partial reachability hide legitimate problems in RIPE's DNSmon
  (\autoref{sec:dnsmon}).
They also explain previously contradictory results in
  several outage detection systems.

\textbf{Artifacts and ethics:}
All of the data used  (\autoref{sec:data_sources})
  and
    created~\cite{ANT22b}
  in this paper
  is available at no cost.
Our work poses no ethical concerns for several reasons (see also \autoref{sec:research_ethics}):
  we collect no additional data, but instead reanalyze existing data
  with new algorithms.
We have no data about individuals, and we do not have or use external
  information to map IP addresses to individuals.
Our work was IRB reviewed and declared non-human subjects research
  (USC IRB IIR00001648).

\section{Partial Reachability and its Outcomes}
	\label{sec:problem}

We next define what we mean by operational reachability
  and how that results in outages, islands, and peninsulas.

\subsection{An Operational Definition of Partial Reachability}
	\label{sec:definition}

Partial reachability
  has a simple operational definition:
  given observations from Vantage Points A and B of destination C
  (\autoref{fig:term_concept}, where reachability is shown by solid lines),
  partial reachability exists if
  A can reach C and B cannot.

This operational definition requires three caveats.
First, problems near the VPs are not very interesting.
(The Internet is not down because my laptop's wifi is off!)
We therefore typically consider partial reachability
  from multiple, independent observers.

This operational definition is tied to specific observers.
Ideally we would like a conceptual definition that does not depend
  on specific observers.

Finally, reachability is defined between ASes,
  but we would like to consider reachability to \emph{the Internet} as a whole.
Here we consider reachability to the \emph{Internet Core};
  the public Internet informally defined
  as what is reachable from today's Tier-1 ISPs.

We believe that addressing these caveats with
  a formal definition, independent of specific observers,
  is an important goal that can lead
  to a formal definition of the Internet Core.
Exploring these questions in depth is outside the scope of this paper,
  but we provide some early results
  elsewhere~\cite{Baltra24a}.
Fortunately,  the above operational definition of reachability
  and informal definition of Internet Core
  are sufficient to develop our algorithms here.

To understand different types of partial reachability,
  we next define outages, islands, and peninsulas.
In \autoref{fig:term_concept},
  in addition to solid lines showing current reachability,
  we also consider dotted lines showing previous reachability.
Each white area is a strongly connected component.
We assume all locations use public Internet addresses
  and so are potentially part of the Internet.

\subsection{Outages}
\label{sec:outages}

In \autoref{fig:term_concept},
  region $X$ shows an \emph{outage},
  where all computers have failed.

\oldreviewfix{S21B7}
A number of groups have examined Internet outages~\cite{Schulman11a,quan2013trinocular,richter2018advancing,guillot2019internet}.
These systems observe the public IPv4 Internet and identify networks
  that are no longer reachable---they have left the Internet.
Often these systems define outages operationally
  (network $b$ is out because none of our \acp{VP} can reach it).
\oldreviewfix{S21B7}
In this paper, we define an outage as when all computers
  in a block are off,
  perhaps due to power loss.
We next define islands,
  when the computers are on but cannot reach the Internet core.

\subsection{Islands: Isolated Networks}
   \label{sec:island}

An \emph{island} is a group of public IP addresses
  partitioned from the Internet,
  but still able to communicate among themselves.
In \autoref{fig:term_concept}
  $D$ and $E$ are islands,
  and $D$ used to be reachable from elsewhere,
  while $E$ has never been reachable.

Operationally, outages and islands are both unreachable from an external \ac{VP},
  but computers in an island are on and can reach each other.

Islands occur when an organization loses
  all connection to the Internet core.
A single-office business with one ISP becomes an island
  when its router's upstream connection fails,
  but computers in the office can still reach each other and in-office servers.
An \emph{address island} is when
  a computer can ping only itself.
Externally, islands and outages appear identical.

\textbf{An Example Island:}
\autoref{fig:a28all_417bca00_accum} shows an example of an island we have observed.
Each line in this graph shows the number of active addresses that are
  estimated from one of 6 observers.
Because Trinocular probes only a few addresses each round,
  these estimates lag the true value
  after an abrupt change in reachability.

The island starts at
  2017-06-03t23:06Z
  and is indicated by the red bar in the middle of the graph.
We see that  \ac{VP} E continues to see all addresses
  because it is inside the island.
By contrast, the other 5 \acp{VP}
  eventually learn they cannot reach any addresses,
  then rediscover addresses after the island ends.
We know this event is an island because we
  confirm \ac{VP} is active, but the other \acp{VP} could not reach it.
Although this brief example is only one /24 block,
  we also see country-sized islands (\autoref{sec:country_sized_islands}).

\subsection{Peninsulas: Partial Connectivity}
	\label{sec:peninsula_definition}

Link and power failures create islands and outages,
  but \emph{partial} connectivity
  is  a more pernicious problem:
  when one can reach some destinations,
  but not others.
\oldreviewfix{S22B16}
We call a group of public IP addresses with partial connectivity
  to other parts of the Internet
  a \emph{peninsula}.
In \autoref{fig:term_concept},
  $B$ and $C$ are on \emph{peninsulas}
  because they cannot directly route to each other,
  although they could relay through $A$.
(Analogous to a geographic peninsula,
  where the mainland may be visible over water,
  but one must detour to reach it,
  so too Internet peninsulas $B$ and $C$ can only reach each other by
  relaying through $A$.)

\oldreviewfix{S22B39}
Peninsulas occur when
  an AS peers with others,
  but lacks routes to parts of the network.
Long-lasting peninsulas often occur due to peering disputes,
  where two ASes refuse to exchange routes~\cite{ipv6peeringdisputes},
  or when an AS purchases transit from a provider
  that is a peninsula.
Peninsula existence has long been recognized,
  with overlay networks designed to route around them~\cite{andersen2001resilient,katz2008studying,katz2012lifeguard}.

\oldreviewfix{S21B2}
\oldreviewfix{S21B4}
\textbf{Examples in IPv6:}
An example of a persistent peninsula is
  the IPv6 peering dispute between Hurricane Electric (HE) and Cogent.
These ISPs decline to peer in IPv6,
  nor are they willing to forward their IPv6 traffic through
  another party.
This problem was noted in 2009~\cite{ipv6peeringdisputes} and
  is visible as of November 2024 in DNSMon~\cite{dnsmon} (\autoref{sec:dnsmon}).
We confirm unreachability
  between HE and Cogent users in IPv6 with traceroutes
  from looking glasses~\cite{he_looking_glass,cogent_looking_glass}
  (HE at 2001:470:20::2 and Cogent at 2001:550:1:a::d).
Neither can reach their neighbor's server,
  but both reach their own.
(Their IPv4 reachability is fine.)

Other IPv6 disputes are Cogent with Google~\cite{google_cogent}, and
 Cloudflare with Hurricane Electric~\cite{cloudflare_he}.
Disputes are often due to an inability to  agree to %
  settlement-free or paid peering.

\textbf{An Example in IPv4:}
We next explore a real-world example of partial reachability to
  several Polish ISPs.
Our algorithms found that
  on 2017-10-23, for a period of 3 hours starting at 22:02Z,
  five Polish \acp{AS}
  had 1716 /24 blocks that were unreachable from five \acp{VP},
  but they remained reachable from a sixth \ac{VP}.

Before the peninsula, all blocks that became partially unreachable
  received service through Multimedia Polska (AS21021, or \emph{MP}),
  via Cogent (AS174), with an alternate path through Tata (AS6453).
\oldreviewfix{S22B41} \oldreviewfix{S22B42}
When the peninsula occurred, traffic continued through Cogent
  but was blackholed; it did not shift to Tata (see \autoref{sec:polish_peninsula_validation}).
One \ac{VP} (W) could reach MP
  through  Tata for the entire event,
  proving MP was connected.
After 3 hours, we see a burst of BGP updates (more than 23k),
  making MP reachable again from all VPs.

\oldreviewfix{S22B43}
To show how our algorithms detect this event (the shaded red region),
  \autoref{fig:a30all_50f5b000_accum} shows how many addresses are
  estimated as reachable.
In this case  \ac{VP} W always reaches the block (W has some green),
  but the others stay unreachable (their estimates fall to zero) for 3 hours.

We can confirm this peninsula with additional observations
  from traceroutes taken by CAIDA's Archipelago~\cite{CAIDA07b} (Ark).
During the event we see 94 unique Ark VPs attempted 345 traceroutes to the affected blocks.
Of the 94 VPs, 21 VPs (22\%) have their last responsive
  traceroute hop in the same \ac{AS} as the
  target address, and 68 probes (73\%) stopped before reaching that \ac{AS}.
The remaining 5 VPs were able to reach the destination \ac{AS} for some
  probes, while not for others.
(Sample traceroutes are in \autoref{sec:polish_peninsula_validation}.)

Although we do not have a root cause for this peninsula from network operators,
  large number of BGP Update messages suggests a routing problem.
 In \autoref{sec:peninsula_locations} we show peninsulas are mostly due to policy choices.

\section{Detecting Partial Connectivity}
	\label{sec:design}

We use observations from multiple, independent \acp{VP}
  to detect partial outages and islands (from \autoref{sec:problem})
  with our new algorithms:
\emph{Taitao} detects peninsulas,
  and \emph{Chiloe}, islands.
(Algorithm names are from Patagonian geography.)

We use these algorithms to study the Internet in
  \autoref{sec:evaluation},
  showing that users see peninsulas as often as outages (\autoref{sec:peninsula_frequency}).
These algorithms and the results help clarify
  prior studies of Internet outages~\cite{Schulman11a,quan2013trinocular,Shah17a,richter2018advancing,guillot2019internet} in \autoref{sec:local_outage_eval}.
We show that they also can improve network observation systems
    such as DNSmon by identifying misconfigurations and persistent problems
    that obscure more urgent, short-term changes
    (\autoref{sec:dnsmon},
    with additional information
  in a preliminary study~\cite{Saluja22a} and detection of Covid-work-from-home~\cite{Song23a}).

\subsection{Suitable Data Sources}
	\label{sec:data_sources}

We evaluate our algorithms
  with publicly available data from several systems:
  USC Trinocular~\cite{quan2013trinocular},
  RIPE Atlas~\cite{Ripe15c},
  and CAIDA's Archipelago~\cite{ark_data},
and use Routeviews~\cite{routeviews} for BGP\@.
\oldreviewfix{S22X10}
We list all datasets in \autoref{tab:datasets} in \autoref{sec:data_sources_list}.

\oldreviewfix{S21A12}
\oldreviewfix{S22A16}
We use data from Trinocular~\cite{quan2013trinocular}
  to study both algorithms
  because it is available at no cost~\cite{LANDER14d},
  provides data since 2014,
  and covers most of the responsive, public IPv4
  Internet~\cite{Baltra20a}.
Briefly, Trinocular watches
  about 5M out of 5.9M responsive IPv4 /24 blocks.
In each probing round of 11 minutes,
  it sends up to 15 ICMP echo-requests (pings),
  stopping early if it proves the block is reachable.
It interprets the results using Bayesian inference,
  and merges the results from six geographically distributed \acp{VP}.
\acp{VP} are in Los Angeles (W), Colorado (C), Tokyo (J), Athens (G),
  Washington, DC (E), and Amsterdam (N).
In \autoref{sec:site_correlation} we show they are topologically independent.
Our algorithms should work with other active probing data
  as future work.

We use
  RIPE Atlas~\cite{Ripe15c}
  to study islands (\autoref{sec:chiloe}),
  and to see how our algorithms
  can improve monitoring (\autoref{sec:dnsmon}).
Atlas consists of about 12k \acp{VP} (as of 2024),
  globally distributed across 3785 different IPv4 ASes.
Atlas VPs carry out both researcher-directed measurements
  and periodic scans of DNS servers.
We use Atlas scans of DNS root servers in our work.

We validate our results using CAIDA's Ark~\cite{ark_data}.
CAIDA ark consisted of about 150 \acp{VP}; each taking traceroutes
  to many IPv4 /24 blocks.

We use BGP data from RouteViews~\cite{routeviews}
  to confirm our data-plane observations.

\oldreviewfix{S21C1}
\oldreviewfix{S21C3}
\oldreviewfix{S21B8}
We generally use recent data, but in some cases we chose older data
  to avoid known problems in measurement systems.
Many of our findings are demonstrated over multiple years (\autoref{sec:2020}).
We use Trinocular measurements for 2017q4 because
  this time period had six active VPs,
  allowing us to make strong statements about multiple perspectives.
It had fewer VPs in 2019 and early 2020,
  but verify and find quantitatively similar results in 2020 in \autoref{sec:2020}.
We use 2020q3 data in
  \autoref{sec:peninsula_locations}
  because Ark observed a very large number of loops in 2017q4.

\subsection{Taitao: a Peninsula Detector}
\label{sec:disagreements}

Peninsulas occur when portions of the Internet %
  are reachable from some locations and not others.
They can be seen by two \acp{VP} disagreeing on reachability.
With multiple VPs, non-unanimous observations
  suggest a peninsula.

Detecting peninsulas presents three challenges.
First, we do not have \acp{VP} everywhere.
\oldreviewfix{S22B47}
If all \acp{VP} are on the same ``side'' of a peninsula ($A$ and $C$ in \autoref{fig:term_concept}),
  their reachability agrees even though \acp{VP} may disagree (like $B$).
\oldreviewfix{S22B48}
Second, observations are often asynchronous.
In Trinocular they are spread over 11~minutes,
  and in Atlas 5~minutes,
  so each VP tests reachability
  at different times.
Observations immediately before and after a network change disagree,
  but both were true when measured---the difference
  is from weak synchronization,
  not a peninsula.
\oldreviewfix{S22B49}
Third, connectivity problems near the observer (or when an observer is an island)
  should not reflect on the intended destination.

We identify peninsulas by detecting disagreements in
  block state by comparing
  valid \ac{VP} observations that occur at about the same time.
Since probing rounds occur every 11 minutes,
  we compare measurements within an 11-minute window.
This approach will see peninsulas that last at least 11 minutes,
  but may miss briefer ones,
  or peninsulas where \acp{VP} are not on ``both sides''.

Formally, $O_{i,b}$ is the set of observers with valid observations
about block $b$ at round $i$.
We look for disagreements in $O_{i,b}$,
  defining $O_{i,b}^\Vs{up} \subset O_{i,b}$ as the
set of observers that measure block $b$ as up at round $i$.
We detect a peninsula when:
\begin{align}
  0 < |O_{i,b}^\Vs{up}| < |O_{i,b}|
\end{align}

When only one \ac{VP} reaches a block,
  that block can be either a peninsula or an island.
We require more information to distinguish them,
  as we describe
 in \autoref{sec:chiloe}.

\subsection{Detecting Country-Level Peninsulas}
\label{sec:detecting_country_peninsulas}

Taitao detects peninsulas based on differences in observations.
Long-lived peninsulas are likely intentional, from policy choices.
One policy is filtering based on national boundaries,
  possibly to implement legal requirements about data sovereignty
  or economic boycotts.

We identify country-specific peninsulas as a special case of Taitao
  where a given destination block is reachable (or unreachable) from only one country,
  persistently for an extended period of time.
(In practice,
  the ability to detect country-level peninsulas is somewhat limited because
  the only country with multiple VPs in our data is the United States.
However, we augment non-U.S.~observers with data from other non-U.S.~sites
  such as Ark or RIPE Atlas.)

A country level peninsula occurs when
  \emph{all} available \acp{VP} from the same country as the target block
  successfully reach the target block and all available \acp{VP} from different countries
  fail.
Formally, we say there is a country peninsula when the set of observers claiming
block $b$ is up at time $i$ is equal to $O_{i,b}^c \subset O_{i,b}$
the set of all available observers with
valid observations at country $c$.
\begin{align}
  O_{i,b}^\Vs{up} = O_{i,b}^c
\end{align}

\subsection{Chiloe: an Island Detector}
	\label{sec:chiloe}

According \autoref{sec:island}, islands occur
when the Internet core is partitioned, and the component
  with fewer than half the active addresses
  is the island.
Typical islands are much smaller.

We can find islands by looking for networks that
  are only reachable from less than half of the Internet core.
However, to classify such networks as an island
  and not merely a peninsula, we need to show that it is partitioned,
  which requires global knowledge.
In addition, if islands are partitioned from all \acp{VP},
  we cannot tell an island,
  with active but disconnected computers,
  from an outage, where they are off.

For these reasons, %
  we must look for islands that include \acp{VP} in their partition.
Because we know the VP is active and scanning
  we can determine how much of the Internet core is in its partition,
  ruling out an outage.
We also can confirm the Internet core is not reachable,
  to rule out a peninsula.

Formally, we say that $B$ is the set of blocks on the Internet core
  responding in the last week.
$B^\Vs{up}_{i,o} \subseteq B$ are blocks reachable from observer
$o$ at round $i$, while
$B^\Vs{dn}_{i,o} \subseteq B$ is its complement.
We detect that observer $o$ is in an island when
  it thinks half or more of the
  observable Internet core
  is down:
\begin{align}
  0 \leq |B^\Vs{up}_{i,o}| \leq |B^\Vs{dn}_{i,o}|
\end{align}
This method is independent of measurement systems,
  but is limited to detecting islands that contain \acp{VP}.
We evaluate islands in Trinocular and Atlas across thousands of VPs in
  \autoref{sec:how_common_are_islands}.
Finally, because observations are not instantaneous,
  we must avoid confusing short-lived islands with long-lived peninsulas.
For islands lasting longer than 11-minutes,
  we also require
 $|B^\Vs{up}_{i,o}| \rightarrow 0$.
With $|B^\Vs{up}_{i,o}| = 0$, it is an address island.

\section{Validating our approach}
	\label{sec:validation}

We validate our algorithms,
  comparing Taitao peninsulas
  and Chiloe islands
  to independent data (\autoref{sec:taitao_validation} and
  \autoref{sec:chiloe_validation}), and examining country-level peninsulas~(\autoref{sec:country_validation}).

\begin{table*}
  \begin{minipage}[b]{.31\linewidth}
  \footnotesize
  \resizebox{\textwidth}{!}{
  \begin{tabular}{c c c | c c c}
    & & & \multicolumn{3}{c}{\normalsize \textbf{Ark}} \\
    & & Sites Up & Conflicting & All Down & All Up \\
	\cline{3-6}
    \multirow{7}{2pt}{\rotatebox[origin=l]{90}{\parbox{1.6cm}{\normalsize \textbf{Trinocular}}}}
    & \multirow{5}{2pt}{\rotatebox[origin=l]{90}{\parbox{35pt}{Conflicting}}}
      & 1 & \cellcolor[HTML]{99ee77}20  & \cellcolor[HTML]{99ee77}6  & \cellcolor[HTML]{FFF9C4}\emph{15}  \\
    & & 2 & \cellcolor[HTML]{99ee77}13  & \cellcolor[HTML]{99ee77}5  & \cellcolor[HTML]{FFF9C4}\emph{11}  \\
    & & 3 & \cellcolor[HTML]{99ee77}13  & \cellcolor[HTML]{99ee77}1  & \cellcolor[HTML]{FFF9C4}\emph{5}   \\
    & & 4 & \cellcolor[HTML]{99ee77}26  & \cellcolor[HTML]{99ee77}4  & \cellcolor[HTML]{FFF9C4}\emph{19}  \\
    & & 5 & \cellcolor[HTML]{99ee77}83  & \cellcolor[HTML]{99ee77}13 & \cellcolor[HTML]{FFF9C4}\emph{201} \\
	\cline{3-6}
    & \multirow{2}{2pt}{\rotatebox[origin=l]{90}{\parbox{21pt}{Agree}}}
    & 0 & \cellcolor[HTML]{F0ABAB}\textbf{6} & \cellcolor[HTML]{CCFF99}97 & \cellcolor[HTML]{F0ABAB}\textbf{6}     \\
    & & 6 & \cellcolor[HTML]{CCFF99}491,120 & \cellcolor[HTML]{CCFF99}90
    & \cellcolor[HTML]{CCFF99}1,485,394 \\
\end{tabular}}
  \caption{Trinocular and Ark agreement table. Dataset A30, 2017q4.}
  \label{tab:taitao_validation_table}
\end{minipage}
\hspace{3mm}
\begin{minipage}[b]{.31\linewidth}
  \centering
  \footnotesize
  \resizebox{\textwidth}{!}{
  \begin{tabular}{c P{40pt} P{40pt} c c}
    & & \multicolumn{3}{c}{\normalsize \textbf{Ark}} \\
    & & Peninsula & \multicolumn{2}{c}{Non Peninsula} \\
    \multirow{2}{2pt}{\rotatebox[origin=l]{90}{\parbox{30pt}{\normalsize \textbf{Taitao}}}}
    & Peninsula & \cellcolor[HTML]{99ee77} 184 & \cellcolor[HTML]{FFF9C4}
    \emph{251 (strict)} & \parbox[30pt][20pt][c]{30pt}{\cellcolor[HTML]{FDD835} \emph{40 (loose)}} \\
    & \parbox[40pt][20pt][c]{40pt}{\centering Non Peninsula} & \cellcolor[HTML]{F0ABAB} \textbf{12} &
    \multicolumn{2}{c}{\cellcolor[HTML]{CCFF99} 1,976,701} \\
  \end{tabular}
}
  \vspace{10pt}
  \caption{Taitao confusion matrix.  Dataset: A30, 2017q4.}
  \label{tab:taitao_confusion_matrix}
\end{minipage}
\hspace{3mm}
  \begin{minipage}[b]{.31\linewidth}
  	\footnotesize
  	\tabcolsep=0.1cm
  	\renewcommand{\arraystretch}{1.1}
\resizebox{\textwidth}{!}{
    \begin{tabular}{c c | c c c | c}
      & & \multicolumn{3}{c}{\normalsize \textbf{Ark}} & \\
      & U.S. VPs & Domestic Only & $\leq5$ Foreign & $>5$ Foreign & Total\\
  	\cline{2-6}
      \multirow{7}{2pt}{\rotatebox[origin=l]{90}{\parbox{50pt}{\centering \normalsize \textbf{Trinocular}}}}
      & WCE & \cellcolor[HTML]{99ee77}211  & \cellcolor[HTML]{99ee77}171  & \cellcolor[HTML]{FFF9C4}\emph{47} & 429  \\
      & WCe & \cellcolor[HTML]{F0ABAB}\textbf{0}  & \cellcolor[HTML]{CCFF99}5  & \cellcolor[HTML]{CCFF99}1 & 6  \\
      & WcE & \cellcolor[HTML]{F0ABAB}\textbf{0}  & \cellcolor[HTML]{CCFF99}1  & \cellcolor[HTML]{CCFF99}0 & 1   \\
      & wCE & \cellcolor[HTML]{F0ABAB}\textbf{0}  & \cellcolor[HTML]{CCFF99}0  & \cellcolor[HTML]{CCFF99}0 & 0  \\
      & Wce & \cellcolor[HTML]{F0ABAB}\textbf{3}  & \cellcolor[HTML]{CCFF99}40 & \cellcolor[HTML]{CCFF99}11 & 54\\
      & wcE & \cellcolor[HTML]{F0ABAB}\textbf{0} & \cellcolor[HTML]{CCFF99}4 & \cellcolor[HTML]{CCFF99}5    & 9     \\
      & wCe & \cellcolor[HTML]{F0ABAB}\textbf{0} & \cellcolor[HTML]{CCFF99}1 & \cellcolor[HTML]{CCFF99}1    & 2 \\
      \midrule
      & Marginal distr.  & 214 & 222 & 65 & 501 \\
  \end{tabular}}
    \caption{Trinocular U.S.-only blocks. Dataset: A30, 2017q4.}
    \label{tab:taitao_countries_validation_table}
  \end{minipage}
\end{table*}

\subsection{Can Taitao Detect Peninsulas?}
	\label{sec:taitao_validation}

\oldreviewfix{S21A15}
We compare Taitao detections
  from 6 \acp{VP}
 to independent observations taken from more than 100 VPs in CAIDA's Ark~\cite{ark_data}.
This comparison is challenging,
  because both Taitao and Ark are imperfect operational systems
  that differ in probing frequency, targets, and method.
\oldreviewfix{S21C7}
Neither defines perfect ground truth,
  but agreement suggests likely truth.

\oldreviewfix{S22B57}
Although Ark probes targets much less frequently than Trinocular,
  Ark makes observations from 171 global locations,
  providing a diverse perspective.
Ark traceroutes also allow us to assess \emph{where} peninsulas begin.
We expect to see a strong correlation between Taitao peninsulas and Ark observations.
\oldreviewfix{S21C8}
(We considered RIPE Atlas as another external dataset,
  but its coverage is sparse, while Ark covers all /24s.)

\textbf{Identifying comparable blocks:}
We study 21 days of Ark observations from 2017-10-10 to -31.
Ark covers all networks with two strategies.
With team probing,
  a 40 VP ``team'' traceroutes to
  all routed /24 about once per day.
For prefix probing,
  about 35 VPs each traceroute to .1 addresses of all routed /24s every day.
We use both types of data: the three Ark teams
  and all available prefix probing VPs.
We group results by /24 block of the traceroute's target address.

Ark differs from Taitao's Trinocular input in three ways:
  the target is a random address or the .1 address in each block;
  it uses traceroute, not ping;
  and it probes blocks daily, not every 11 minutes.
Sometimes these differences cause Ark traceroutes to fail
  when a simple ping succeeds.
\oldreviewfix{S22B58}
First, Trinocular's targets respond more often because
  it uses a curated hitlist~\cite{Fan10a}
  while Ark does not.
Second, Ark's traceroutes can terminate due to path
  \emph{loops}
  or \emph{gaps} in the path,
  (in addition to succeeding or reporting target unreachable).
We do not consider results with gaps, %
  so problems on the path do not
  bias results for endpoints reachable by direct pings.

\oldreviewfix{S22B59} \oldreviewfix{S22B60}
To correct for differences in target addresses,
  we must avoid
  misinterpreting a block as unreachable
  when the block is online but Ark's target address is not,
  we discard traces sent to never-active addresses
  (those not observed in 3 years of complete IPv4 scans), %
  and blocks
  for which Ark did not get a single successful response.%
\oldreviewfix{I23D7, and below}
Since
  dynamic addressing~\cite{Padmanabhan16a} means Ark often
  fails with an unreachable last hop,
  we see conflicting observations in Ark, implying false peninsulas.
We therefore trust Ark confirmation of outages and full reachability,
  but question Ark-only peninsulas.

\oldreviewfix{S21A14}
To correct for Ark's less frequent probing,
  we compare \emph{long-lived} Trinocular down-events (5 hours or more).
Ark measurements are infrequent (once every 24 hours) compared to Trinocular's 11-minute reports,
  so short Trinocular events are often unobserved by Ark.
To confirm agreements or conflicting reports from Ark,
  we require at least 3 Ark observations within the peninsula's span of time.

We filter out blocks with frequent transient changes
  or signs of network-level filtering.
We define the ``reliable'' blocks suitable for comparison
  as those responsive for at least 85\% of the quarter
  from each of the 6 Trinocular VPs.
(This threshold avoids diurnal blocks or blocks with long outages;
  values of 90\% or less have similar results.)
We also discard flaky blocks whose responses are frequently inconsistent across \acp{VP}.
(We consider more than 10 combinations of \ac{VP} as frequently inconsistent.)
\oldreviewfix{S22B61}
For the 21 days, we find 4M unique Trinocular /24 blocks,
  and 11M Ark /24 blocks,
  making 2M blocks in both available for study.

\textbf{Results:}
\autoref{tab:taitao_validation_table} provides details
  and \autoref{tab:taitao_confusion_matrix} summarizes our interpretation,
 treating Taitao as prediction and Ark as truth.
Here dark green indicates true positives (TP):
  when (a) either both Taitao and Ark show mixed results, both indicating a peninsula,
  or when (b) Taitao indicates a peninsula (1 to 5 sites up but at least one down),
Ark shows all-down during the event and up before and after.
We treat Ark in case (b) as positive
  because the infrequency of Ark probing
  (one probe per team every 24 hours) %
  means we cannot guarantee
  VPs in the peninsula will probe responsive targets in time.
Since peninsulas are not common, so too are true positives,
  but we see 184 TPs.

We show \emph{true negatives} as light green and neither bold nor italic.
In almost all of these cases (1.4M)
  both Taitao and Ark reach the block, agreeing.
\oldreviewfix{I23D7, and above}
The vast majority of these are an artifact of our use of
  Ark as ``ground truth'', when it is not designed
  to accurately measure partitions.
The challenge an Ark claim of peninsula is that
  about  5/6ths of Ark probes fail in the last hop because it probes a single
  random address (see \cite{quan2013trinocular} figure 6).
As  a result, while positive Ark results support non-partitions,
  negative Ark results are most likely a missed target and not
  an unreachable block.
We therefore treat this second most-common result (491k cases) as a true negative.
While this is likely true 5/6ths of the time,
  it is also likely that the 150 \acp{VP} in Ark
  can see some peninsulas that our 6 \acp{VP} miss---thus
  our results may \emph{underestimate}
  the severity of the problem of partial connectivity.
Our validation therefore demonstrates a strong \emph{lower bound}
  on the number of peninsulas,
  hopefully prompting a tighter bound in future work.
We expand on this analysis and its interpretation in \autoref{sec:false_negative_details}.

\oldreviewfix{I23D20,I23D21}
For the same reason, we include the small number (97) of cases where
  both Ark and Taitao report all-down as true negatives,
  assuming Ark terminates at an empty address.
We include in this category the 90 events where Ark is all-down and Trinocular
is all-up.
We attribute Ark's failure to reach its targets to infrequent probing.

We mark \emph{false negatives} as red and bold.
For these few cases (only 12),
  all Trinocular \acp{VP} are down, but
  Ark reports all or some responding.
We believe these cases indicate blocks that have chosen to drop Trinocular traffic.

\oldreviewfix{S22B63}
Finally, yellow italics shows when Taitao's peninsulas
  are \emph{false positives}, since all Ark probes reached the target block.
This case occurs when either traffic from some Trinocular \acp{VP} is
  filtered, or all Ark VPs are ``inside'' the peninsula.
Light yellow (strict) shows all the 251 cases that Taitao detects.
\oldreviewfix{S22B64}
For most of these cases (201),
  five Trinocular \acp{VP} responding and one does not,
  suggesting network problems are near one of the Trinocular VPs
  (since five of six independent VPs have working paths).
Discarding these cases we get 40 (orange); still conservative but a \emph{looser} estimate.

The strict scenario sees precision 0.42, recall 0.94, and $F_1$ score
  0.58,
  and in the loose scenario, precision improves to 0.82 and
  $F_1$ score to 0.88.
We consider these results a strong lower bound on the size of problem,
  and confirmation that the peninsulas detected by Taitao are correct.
We hope our results will prompt future work to tighten our bound on size.

\subsection{Detecting Country-Level Peninsulas}
	\label{sec:country_validation}

Next, we verify detection of country-level peninsulas
(\autoref{sec:detecting_country_peninsulas}).
We expect that legal requirements sometimes result in long-term
  network unreachability.
For example, blocking access from Europe
  is a crude way to comply with the EU's GDPR~\cite{eu_gdpr}.

Identifying country-level peninsulas requires
  multiple VPs in the same country.
Unfortunately the source data we use only has multiple \acp{VP} for the United States.
We therefore look for U.S.-specific peninsulas
  where only these \acp{VP} can reach the target and the non-U.S.-\acp{VP} cannot,
  or vice versa.

We first consider the 501 cases where Taitao reports that only U.S.~\acp{VP}
  can see the target, and compare to how Ark \acp{VP} respond.
For Ark, we follow
\autoref{sec:taitao_validation},
  except retaining blocks with less than 85\% uptime.
We only consider Ark VPs that are able
  to reach the destination (that halt with ``success'').
We note blocks that can only be reached by Ark VPs within the same
country as domestic, and blocks that can be reached from VPs located in other
countries as foreign.

In \autoref{tab:taitao_countries_validation_table}
  we show the number of blocks that uniquely
  responded to all U.S.~\ac{VP} combinations during the quarter.
We contrast these results against Ark reachability.

True positives are when  Taitao shows a peninsula
  responsive only to U.S.~\acp{VP}
  and nearly all Ark \acp{VP} confirm this result.
We see 211 targets are U.S.-only, and another 171 are available to only a few
  non-U.S.~countries.
The specific combinations vary: sometimes allowing access from the U.K.,
  or Mexico and Canada.
Together these make 382 true positives, most of the 501 cases.
\oldreviewfix{S22X9}
Comparing all positive cases, we see a
  very high precision of 0.99 (382 green of 385 green and red reports)---our
  predictions are nearly all confirmed by Ark.

In yellow italics we show 47 cases of false positives
  where more than five non-U.S. countries are allowed access.
In many cases these include many European countries.
\oldreviewfix{S22X9}
Our recall is therefore 0.89 (382 green of 429 green and yellow true country peninsulas).

In light green we show true negatives.
Here we include blocks that filter one or more U.S. \acp{VP},
and are reachable from Ark VPs in multiple
countries, amounting to a total of 69 blocks.
There are other categories involving non-U.S. sites,
  along with other millions of true
  negatives, however, we only concentrate in these few.

In red and bold we show three false negatives.
These three blocks seem to have strict filtering policies,
  since they were reachable only from one U.S.~site (W)
  and not the others (C and E) in the 21 days period.

\subsection{Can Chiloe Detect Islands?}
\label{sec:chiloe_validation}

Chiloe (\autoref{sec:chiloe}) detects islands when a \ac{VP} within the island
  can reach less than half the rest of the world.
When less than 50\% of the network replies,
  it means that the \ac{VP} is either in an island (for brief events, or when replies drop near zero)
  or a peninsula (long-lived partial replies).

To validate Chiloe's correctness,
  we compare when a single \ac{VP} believes to be in an island,
  against what the rest of the world believes about that \ac{VP}.

\oldreviewfix{S22B69}
\oldreviewfix{I23B5}
Islands are unreachable, like $D$ in \autoref{fig:term_concept}.
We measure blocks,
  so if any address in block $D$ can reach another, it is an island.
If no external \acp{VP} can reach $D$'s block,
  Chiloe confirms an island,
  but some \ac{VP} reaching $D$'s block implies a peninsula.
In \autoref{sec:site_correlation} we show that Trinocular \acp{VP} are independent,
  and therefore no two \acp{VP} live within the same island.
We believe this definition is the best possible ground truth,
  since perfect classification requires instant,
  global knowledge and cannot be measured in practice.

We take 3 years worth of data from all six
  Trinocular \acp{VP}.
\oldreviewfix{S22B68}
From Trinocular's pacing, we analyze 11-minute bins.

  \begin{table}
\centering
	\resizebox{0.49\columnwidth}{!}{
\subfloat[Chiloe confusion matrix \label{tab:chiloe_validation}]{%
    \centering
    \begin{tabular}{c c P{30pt} P{35pt} @{}m{0pt}@{}}
      & & \multicolumn{2}{c}{\normalsize \textbf{Chiloe}} \\
      & & \parbox{40pt}{\centering Island} & \parbox{40pt}{\centering Peninsula} \\
      \multirow{3}{10pt}{\rotatebox[origin=l]{90}{\parbox{50pt}{\normalsize \textbf{Trinocular}}}} &
      \parbox[10pt][20pt][c]{40pt}{Blk Island} &
      \cellcolor[HTML]{99ee77} 2 & \cellcolor[HTML]{F0ABAB} \textbf{0} \\
      & \parbox[10pt][20pt][c]{50pt}{Addr Island} &
      \cellcolor[HTML]{99ee77} 19 & \cellcolor[HTML]{F0ABAB} \textbf{8} \\
      & \parbox[10pt][20pt][c]{40pt}{Peninsula}
      & \cellcolor[HTML]{FFF9C4} \emph{2} & \cellcolor[HTML]{CCFF99} 566
    \end{tabular}

	  }}%
\qquad%
	\resizebox{0.42\columnwidth}{!}{
\subfloat[Detected islands\label{tab:island_summary}]{%
    \centering
  \begin{tabular}{c c c}
   Sites	& Events	& \oldreviewfix{S22B87}Per Year \\
   \midrule
   W	    & 5		    & 1.67 \\
   C	    & 11  		& 3.67 \\
   J	    & 1		    & 0.33 \\
   G	    & 1		    & 0.33 \\
   E	    & 3		    & 1.00 \\
   N	    & 2		    & 0.67 \\
    \hline
    All (norm.)     & 23        & 7.67 (1.28)  \\
  \end{tabular}

	  }}
    \caption{(a) Chiloe confusion matrix, events between 2017-01-04 and 2020-03-31, datasets A28 through A39.
(b) Islands detected from 2017q2 to 2020q1.
	  }
\vspace*{-2ex}
\end{table}

In \Cref{tab:chiloe_validation} we show that Chiloe detects 23 islands
across three years.
\oldreviewfix{I23D13}
In 2 of these events, the block is unreachable from other \acp{VP},
  confirming the island with our validation methodology.
Manual inspection confirms that the remaining
19 events are islands too, but at the address level---the \ac{VP}
  was unable to reach anything but did not lose power,
  and other addresses in its block were reachable from \acp{VP} at other locations.
\oldreviewfix{S22B70}
These observations suggest a VP-specific problem making it an island.
Finally, for 2 events, the prober's block was reachable during the event by
every site including the prober itself which suggests partial connectivity
(a peninsula), and therefore a false positive.

In the 566 non-island events (true negatives),
  a single \ac{VP} cannot reach more than 5\% but less than 50\% of
  the Internet core.
In each of these cases, one or more
other \acp{VP} were able to reach the affected \ac{VP}'s block,
  showing they were not an island (although perhaps a peninsula).
The table omits the frequent events when less than 5\% of the network is unavailable from the \ac{VP},
  although they too are true negatives.

Bold red shows 8 false negatives. These are events that last about 2 Trinocular
rounds or less (22 min), often not enough time for Trinocular to change its
belief on block state.

\subsection{Are the Sites Independent?}
	\label{sec:site_correlation}

Our evaluation assumes \acp{VP} do not share common network paths.
Two \acp{VP} in the same location would share the same local outages,
  but those in different physical locations
  will often use different network paths,
  particularly with today's ``flatter'' Internet~\cite{Labovitz10c}.
We next quantify this similarity to validate our assumption.

We next measure similarity of observations
  between pairs of VPs.
We examine only cases where one of the pair disagrees with some other VP,
  since when all agree, we have no new information.
If the pair agrees with each other, but not with the majority,
  the pair shows similarity.
If they disagree with each other, they are dissimilar.
We quantify similarity $S_P$ for a pair of sites $P$ as
	${S_P = (P_1 + P_0)/(P_1 + P_0 + D_*)}$,
where $P_s$ indicates the pair agrees on the network having state $s$ of
  up (1) or down (0) and disagrees with the others,
  and for $D_*$, the pair disagrees with each other.
$S_P$ ranges from 1, where the pair always agrees,
  to 0, where they always disagree.

\begin{table*}

\raisebox{4mm}{\begin{minipage}[b]{.24\textwidth}
  \resizebox{\textwidth}{!}{
    \footnotesize
	\begin{tabular}{c|c@{\hspace{0.7ex}}c@{\hspace{0.7ex}}c@{\hspace{0.7ex}}c@{\hspace{0.7ex}}c}
       & C      & J      & G      & E      & N      \\
	\hline
	W  & 0.017  & 0.031  & 0.019  & 0.035  & 0.020  \\
    C  &        & 0.077  & 0.143  & 0.067  & 0.049  \\
    J  &        &        & 0.044  & 0.036  & 0.046  \\
    G  &        &        &        & 0.050  & 0.100  \\
    E  &        &        &        &        & 0.058  \\
    \end{tabular}
    }
    \captionsetup{type=table}
    \caption{Similarities all VPs. Dataset: A30, 2017q4.}
    \label{tab:overall_correlation}
\end{minipage}}
\hspace{3ex}
  \begin{minipage}[b]{.31\linewidth}
  \resizebox{\textwidth}{!}{
    \footnotesize
    \begin{tabular}{c r r | r r}
      & \multicolumn{2}{c | }{\textbf{Target AS}}
      & \multicolumn{2}{c}{\textbf{Target Prefix}} \\
      Sites Up & At & Before & At & Before \\
      \midrule
      0	& 21,765	    & 32,489	    & 1,775	    & 52,479 \\
      \rowcolor[HTML]{DCDCDC}
      1	& 587	    & 1,197	    & 113	    & 1,671 \\
      \rowcolor[HTML]{DCDCDC}
      2	& 2,981	    & 4,199	    & 316	    & 6,864 \\
      \rowcolor[HTML]{DCDCDC}
      3	& 12,709	    & 11,802	    & 2,454	    & 22,057 \\
      \rowcolor[HTML]{DCDCDC}
      4	& 117,377	& 62,881	    & 31,211	    & 149,047 \\
      \rowcolor[HTML]{DCDCDC}
      5	& 101,516	& 53,649	    & 27,298	    & 127,867 \\
  	\cline{2-5}
      \rowcolor[HTML]{DCDCDC}
      \textbf{1-5} & \cellcolor[HTML]{99ee77} \textbf{235,170} & \textbf{133,728} & \textbf{61,392} & \cellcolor[HTML]{99ee77}\textbf{307,506} \\
      6	& 967,888	& 812,430	& 238,182	& 1,542,136 \\
    \end{tabular}}
    \caption{Halt location of failed traceroutes for peninsulas longer than 5
    hours. Dataset A41, 2020q3.}
    \label{tab:peninsula_root_cause}
  \end{minipage}
\hspace{3ex}
  \begin{minipage}[b]{.24\linewidth}
  \resizebox{\textwidth}{!}{
    \footnotesize
    \begin{tabular}{l c c}
            Industry           & ASes  & Blocks   \\
            \midrule
            ISP               & 23  & 138 \\
            Education         & 21  & 167 \\
            Communications    & 14  & 44  \\
            Healthcare        & 8   & 18  \\
            Government        & 7   & 31  \\
            Datacenter        & 6   & 11  \\
            IT Services       & 6   & 8   \\
            Finance           & 4   & 6   \\
            Other (6 types)
            & \multicolumn{2}{c}{6 (1 per type)}   \\
    \end{tabular}}
    \caption{U.S. only blocks. Dataset A30, 2017q4}
    \label{tab:industry}
  \end{minipage}
\end{table*}

\autoref{tab:overall_correlation} shows similarities for each pair
of the 6 Trinocular VPs (as half of the symmetric matrix).
No two sites have a similarity more than 0.14,
  and most pairs are under 0.08.
This result shows that no two sites are particularly correlated.

\section{Internet Islands and Peninsulas}
	\label{sec:evaluation}

We now examine islands and peninsulas in the Internet core.

\subsection{How Common Are Peninsulas?}
	\label{sec:peninsula_frequency}

\oldreviewfix{S22X3: clearer statement about result}
We estimate how often peninsulas occur  in the Internet core
  in three ways.
First, we directly measure the visibility of peninsulas %
  by summing the duration of peninsulas as seen from six VPs.
Second, we confirm the accuracy of this estimate
  by evaluating its convergence as we vary the number of VPs---more VPs
  show more peninsula-time, but a result that converges
  suggests it is
  approaching the limit.
Third, we compare peninsula-time to outage-time,
  showing that, in the limit, observers see both for about the same
  duration.
\oldreviewfix{S21B9}
Outages correspond to service downtime~\cite{down_time_cost},
  and are a recognized problem in academia and industry.
Our results show that \emph{peninsulas are as common as outages},
  suggesting peninsulas are an important new problem deserving attention.

\textbf{Peninsula-time:}
We estimate the duration an observer can see a peninsula
  by considering three types of events: \emph{all up}, \emph{all down}, and
  \emph{disagreement} between six VPs.
Disagreement, the last case, suggests a peninsula,
  while agreement (all up or down),
  suggests no problem or an outage.
We compute peninsula-time by summing the time each target /24
  has disagreeing observations from Trinocular VPs.

\begin{figure*}
\adjustbox{valign=b}{\begin{minipage}[b]{.28\linewidth}
    \includegraphics[width=1\linewidth]{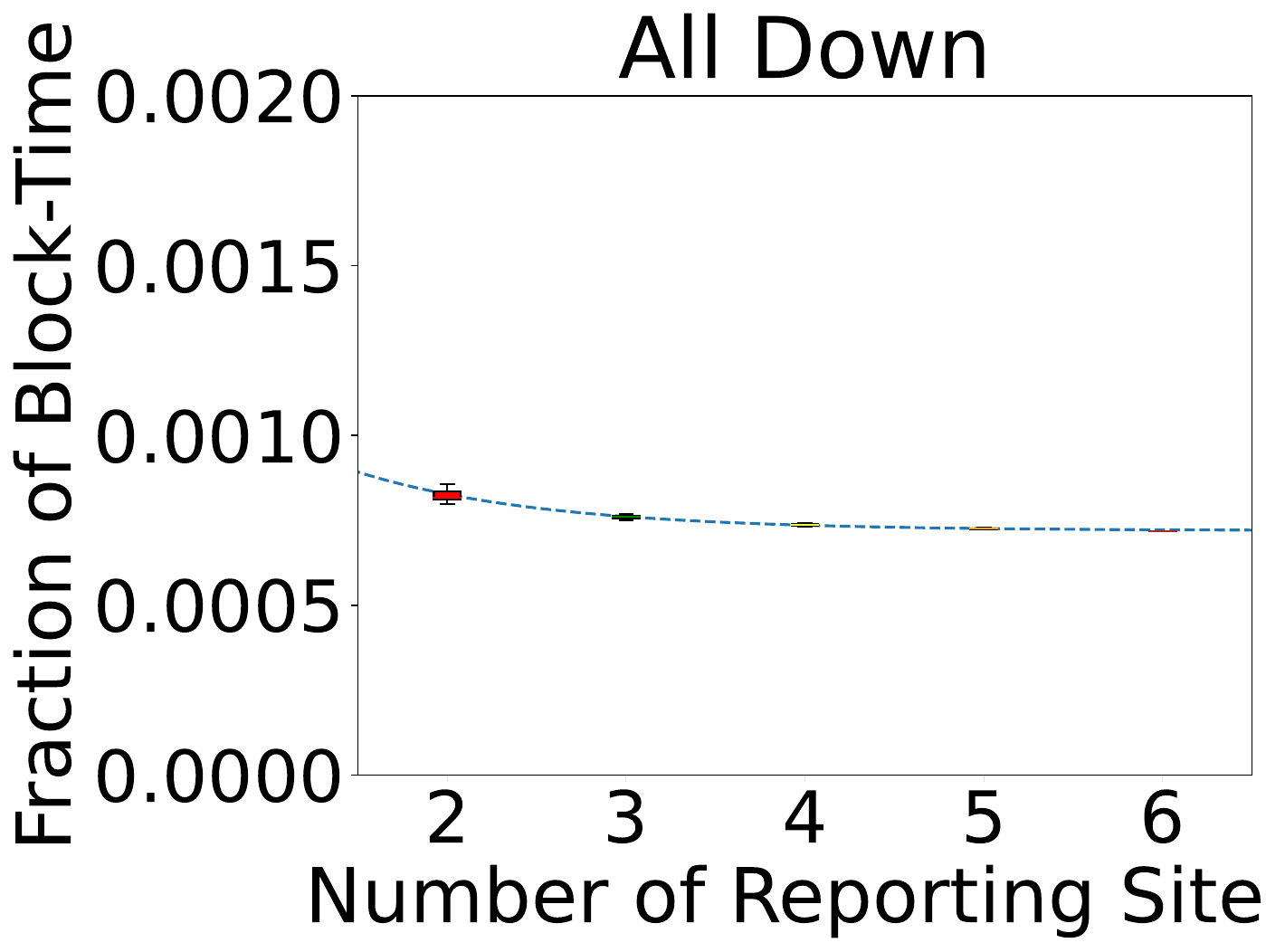}
\end{minipage}}\quad
\adjustbox{valign=b}{\begin{minipage}[b]{.28\linewidth}
    \includegraphics[width=1\linewidth]{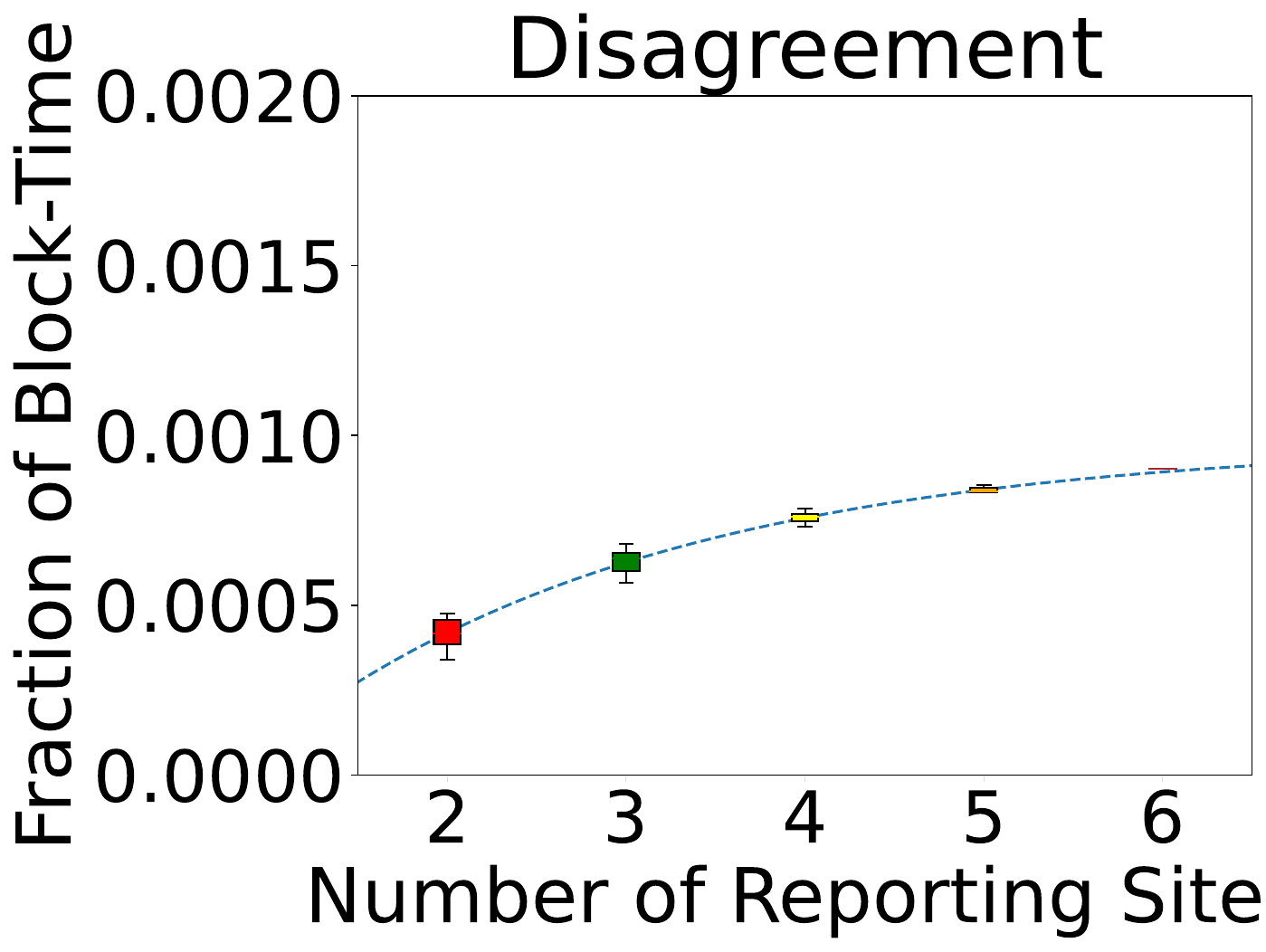}
\end{minipage}}\quad
\adjustbox{valign=b}{\begin{minipage}[b]{.37\linewidth}
    \includegraphics[width=.95\linewidth]{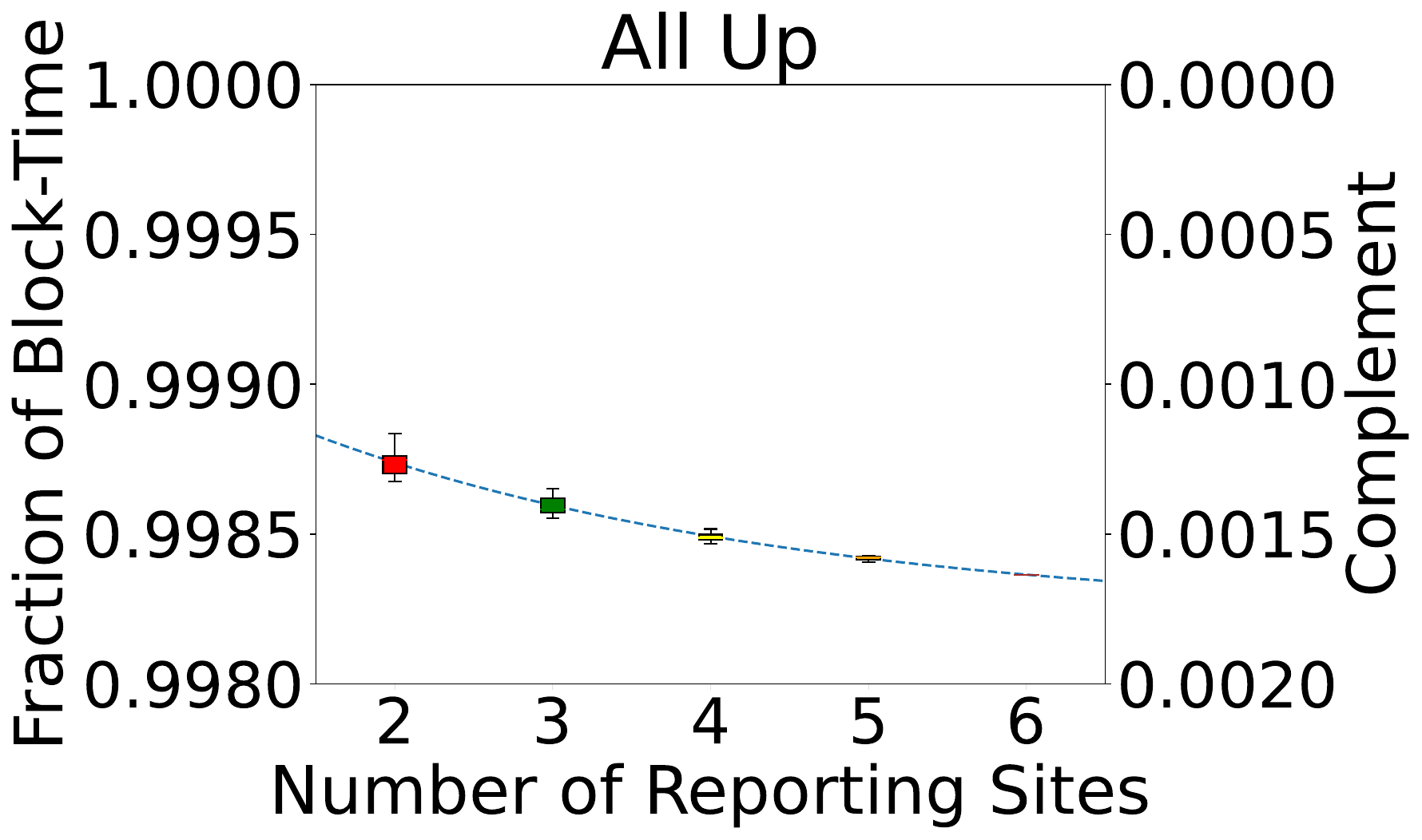}
\end{minipage}}\quad
\caption{Distribution of block-time fraction: all-down
        (left), disagreement (center), and all-up (right), events $\ge 1$ hour.
        Data: 3.7M blocks, 2017-10-06 to -11-16, A30.}
\label{fig:a30all_peninsulas_duration_oct_nov}
\end{figure*}

We have computed peninsula-time
  by evaluating Taitao over Trinocular data for 2017q4~\cite{LANDER14d}.
\autoref{fig:a30all_peninsulas_duration_oct_nov} shows the distribution of peninsulas
measured as a fraction of block-time for an increasing number of sites.
We consider all possible combinations of the six sites.

First we examine the data with all 6 VPs---the rightmost point on each graph.
We see that %
  peninsulas (the middle, disagreement graph)
  are visible about 0.00075 of the time.
This data suggests \emph{peninsulas regularly occur,
  appearing at least 0.1\% of the time}.
Fortunately, large peninsulas are rare from many locations---our 6 VPs almost always see the same targets.

\textbf{Convergence:}
With more \acp{VP} we get a better view of the Internet core's overall state.
As more reporting sites are added, more peninsulas are discovered.
\oldreviewfix{S22B75}
That is previously inferred outages (all unreachable) should have been peninsulas,
  with partial reachability.
All-down (left) decreases from an average of 0.00082 with 2 \acp{VP} to 0.00074 for 6
\acp{VP}. All-up (right) goes down a relative 47\% from 0.9988 to 0.9984, while
disagreements (center) increase from 0.0029 to 0.00045.
Outages (left) converge after 3 sites,
  as shown by the fitted curve and decreasing variance.
Peninsulas and all-up converge more slowly.
We conclude that \emph{a few, independent sites (3 or 4) converge on a good estimate of true islands
  and peninsulas}.

We support this claim by comparing
  all non-overlapping combinations of 3 sites.
If all combinations are equivalent,
  then a fourth site will not add new information.
Six \acp{VP} yield 10 possible sets of 3 sites;
  we examine those combinations for each of 21 quarters, from 2017q2 to 2020q1.
When we compare the one-sample Student $t$-test
  to evaluate if the difference of each pair of combinations of those 21 quarters
  is greater than zero.
None of the combinations are rejected at confidence level 99.75\%,
  suggesting that any combination of three sites is statistically equivalent
  and confirm our claim that a few sites are sufficient for estimation.

\textbf{Relative impact:}
Finally, comparing outages (the left graph) with peninsulas (the middle graph),
  we see both occur about the same fraction of time (around 0.00075).
This comparison shows that \emph{peninsulas are about as common as outages},
  suggesting they deserve more attention.

\textbf{Generalizing:}
We confirm that each of these results holds in a subsequent year
  in \autoref{sec:2020},
  suggesting the result is not unique to this quarter.
While we reach a slightly different limit (in that case,
  peninsulas and outages appear about in 0.002 of data),
  we still see good convergence after 4 VPs.

\subsection{How Long Do Peninsulas Last?}
	\label{sec:peninsula_duration}

\oldreviewfix{S21A16}
Peninsulas have multiple root causes:
  some are short-lived routing misconfigurations
  while others reflect long-term disagreements in routing policy.
In this section we study the distribution of peninsulas in terms of their duration
  to determine the prevalence of persistent peninsulas.
\oldreviewfix{S21A16}
\oldreviewfix{S22B78}
We will show that there are millions of brief peninsulas,
  likely due to transient routing problems,
  but that 90\% of peninsula-time is in long-lived events (5\,h or more, following~\autoref{sec:taitao_validation}).

We use Taitao to see peninsula duration for all detected in 2017q4:
   some 23.6M peninsulas affecting 3.8M unique blocks.
If instead we look at \emph{long-lived} peninsulas (at least 5\,h),
  we see 4.5M peninsulas in 338k unique blocks.

\oldreviewfix{S22B79}
\autoref{fig:a30_partial_outages_duration_cdf} examines peninsula
  duration
  in three ways:
  a cumulative distribution (CDF) counting all peninsula events
  (left, solid, purple line),
  the CDF of the number of peninsulas for VP-down events
  longer than 5 hours (middle, solid green line),
  and the cumulative size of peninsulas
  for VP down events longer than 5 hours (right, green dashes).

\oldreviewfix{S22X7}
We see that there are many very brief peninsulas (purple line):
  about 65\% last only 20--60 minutes ($\sim$2--6 observations).
With two or more observations, these events are not just
  one-off measurement loss.
These results suggest that while the Internet core is robust,
there are many small connectivity glitches (7.8M events).
Events that are two rounds (20 minutes) or shorter
  may be due to transient BGP blackholes~\cite{bush2009internet}.

The number of day-long or multi-day peninsulas is small,
  only 1.7M events (2\%, the purple line).
However, about 57\% of all peninsula-time is in such longer-lived events
  (the right, dashed line),
  and 20\% of time is in events lasting 10 days or more,
  even when longer than 5 hours events are less numerous (compare the middle, green line to the left, purple line).
\oldreviewfix{S22B81}
Events lasting a day last long enough that they can be debugged by human network operators,
  and events lasting longer than a week suggest potential
  policy disputes and \emph{intentional} unreachability.
Together, these long-lived events suggest that
  there is benefit to identifying non-transient peninsulas
  and addressing the underlying routing problem.

\subsection{Where Do Peninsulas Occur?}
	\label{sec:peninsula_locations}

Firewalls, link failures, and routing problems cause peninsulas on the Internet.
These can either occur inside a given AS,
  or in upstream providers.

To detect where the Internet breaks into peninsulas,
  we look at traceroutes that failed to reach their target address,
  either due to a loop or an ICMP unreachable message.
Then, we find where these traces halt, and
  take note whether halting occurs \emph{at} the target AS and target prefix,
  or \emph{before} the target AS and target prefix.

For our experiment
  we run Taitao to detect peninsulas at target blocks over Trinocular VPs,
  we use Ark's traceroutes~\cite{ark_data_2020} to find last IP address before halt, and
  we get target and halting ASNs and prefixes using RouteViews.

In~\autoref{tab:peninsula_root_cause} we show how many traces halt
\emph{at} or \emph{before} the target network.
The center, gray rows show peninsulas (disagreement between \acp{VP})
  with their total sum in bold.
For all peninsulas (the bold row),
  more traceroutes halt at or inside the target AS (235k vs.~134k, the left columns),
  but they more often terminate before reaching the target prefix (308k vs.~61k, the right columns).
This difference suggests policy is implemented at or inside ASes, but not at routable prefixes.
By contrast, outages (agreement with 0 sites up)
  more often terminate before reaching the target AS.
Because peninsulas are more often at or in an AS,
  while outages occur in many places,
  it suggests that peninsulas are policy choices.

\subsection{What Sizes Are Peninsulas?}
	\label{sec:peninsula_size}

When network issues cause connectivity problems like peninsulas,
  the \emph{size} of those problems may vary,
  from country-size%
  \, (see \autoref{sec:country_peninsulas})%
, to \ac{AS}-size,
and also for routable prefixes or fractions of prefixes.
We next examine peninsula sizes.

We begin with Taitao peninsula detection at a /24 block level.
We match peninsulas across blocks within the same prefix by start time and
duration, both measured in one hour timebins.
This match implies that the Trinocular \acp{VP} observing the blocks as up are
also the same.

We compare peninsulas to routable prefixes from Routeviews \cite{routeviews},
  using
  longest prefix matches with /24 blocks.

Routable prefixes consist of many blocks, some of which may not be measurable.
We therefore define the \emph{peninsula-prefix fraction}
  for each routed prefix as fraction of blocks in the peninsula
  that are Trinocular-measurable blocks.
To reduce noise provided by single block peninsulas,
  we only consider peninsulas covering 2 or more blocks in a prefix.

\begin{figure*}
  \begin{minipage}[b]{.6\linewidth}
  \subfloat[Number of Peninsulas]{
    \includegraphics[width=0.48\linewidth]{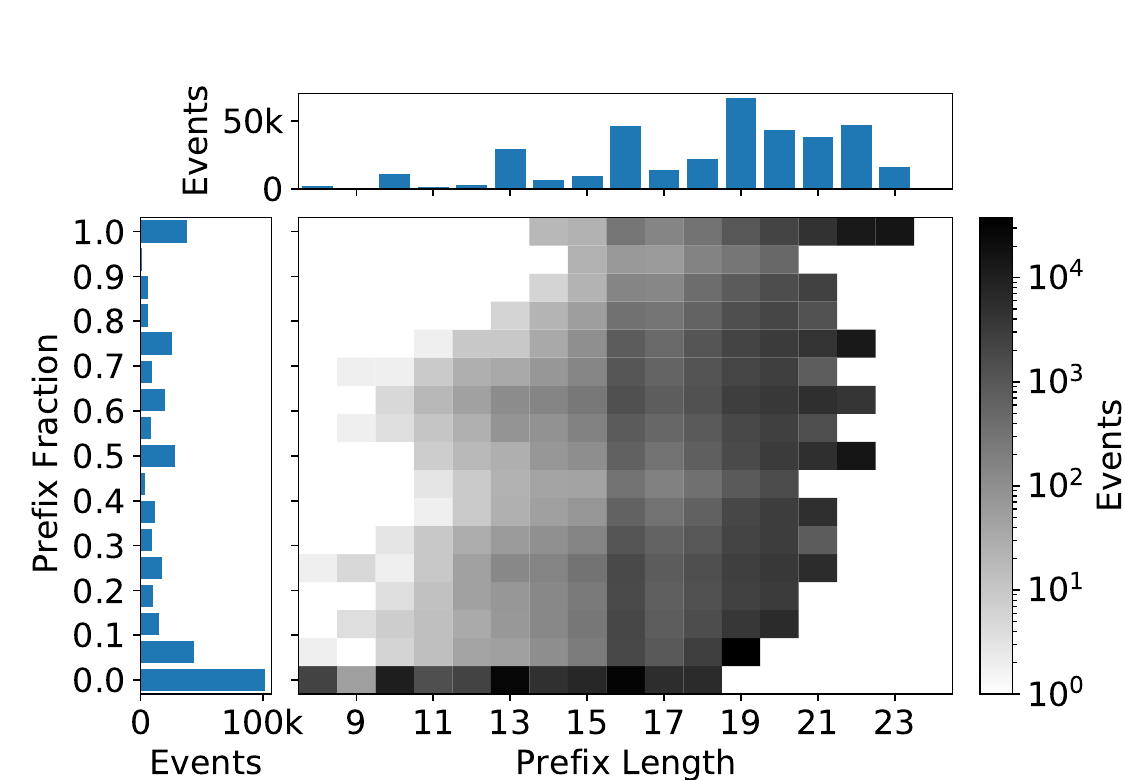}
    \label{fig:a30all_blocks_in_prefix_prefix_fraction_heatmap}
  }
\quad
  \subfloat[Duration fraction]{
    \includegraphics[width=0.48\linewidth]{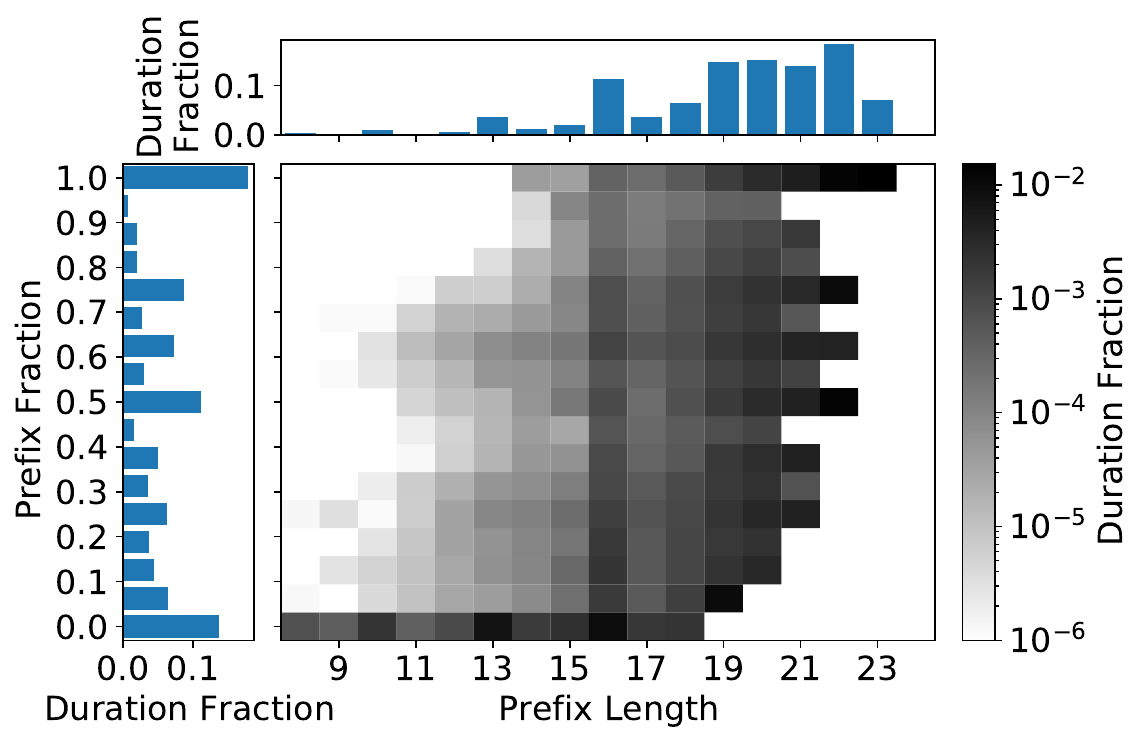}
    \label{fig:a30all_blocks_in_prefix_prefix_fraction_heatmap_duration}
  }
  \caption{Peninsulas measured with per-site down events longer than 5 hours. Dataset A30, 2017q4.}
  \end{minipage}
\quad
\begin{minipage}[b]{.3\linewidth}
        \includegraphics[width=1\linewidth]{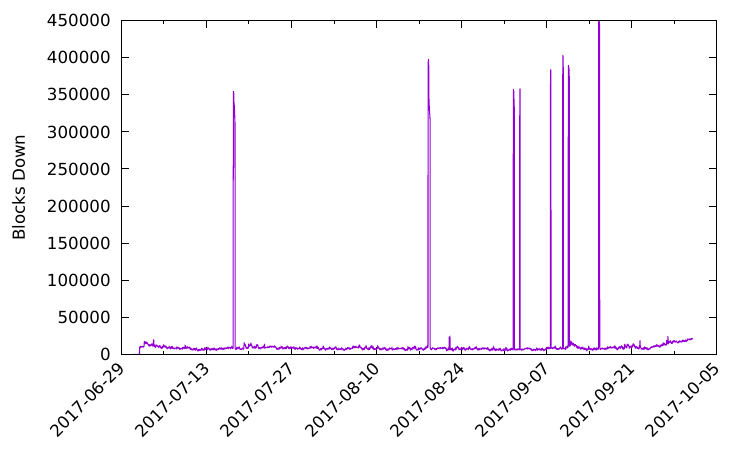}
	\caption{Unreachable blocks over time.  Large spikes are unreachability to Chinese-allocated IPv4 addresses. Dataset: A29, 2017q3.}
  \label{fig:a29all_outagedownup_4096}

\end{minipage}
\end{figure*}

\autoref{fig:a30all_blocks_in_prefix_prefix_fraction_heatmap} shows the number
of peninsulas for different prefix lengths and the fraction of the prefix
affected by the peninsula
  as a heat-map,
  where we group them into bins.

We see that about 10\% of peninsulas
  are likely due to
  routing problems or policies,
  since 40k peninsulas affect the whole routable prefix.
However, a third of peninsulas
  (101k, at the bottom of the plot)
  affect
  only a very small fraction of the prefix.
\oldreviewfix{S21C11}
These low prefix-fraction peninsulas suggest
  that they happen \emph{inside} an ISP and
  are not due to interdomain routing.

Finally, we show that \emph{longer-lived peninsulas are likely due to routing or policy choices}.
\autoref{fig:a30all_blocks_in_prefix_prefix_fraction_heatmap_duration}
  shows the same data source,
  but weighted by fraction of time each peninsula
  contributes to the total peninsula time during 2017q4.
Here the larger fraction of weight are peninsulas covering
  full routable prefixes---20\% of all peninsula time during the
quarter (see left margin).

\subsection{Country-Level Peninsulas}
	\label{sec:country_peninsulas}

Country-specific filtering is a routing policy made by networks to
restrict traffic they receive.
We next look into  what type of organizations actively block overseas
traffic.
For example, good candidates to restrain who can reach them for security purposes
  are government related organizations.

We test for country-specific filtering (\autoref{sec:detecting_country_peninsulas}) over 2017q4 and find 429
unique U.S.-only blocks in 95 distinct ASes.
We then manually verify each AS categorized by industry
   in \autoref{tab:industry}.
It is surprising how many universities filter by country.
While not common, country specific blocks do occur.

\subsection{How Common Are Islands?}
	\label{sec:how_common_are_islands}

Multiple groups have shown that there are many network outages in the Internet~\cite{Schulman11a,quan2013trinocular,Shah17a,richter2018advancing,guillot2019internet}.
\oldreviewfix{S22B86}
We have described (\autoref{sec:problem}) two kinds of outages:
  full outages where all computers at a site are down (perhaps due to a loss of power),
  and islands, where the site is cut off from the Internet core, but computers
    at the site can talk between themselves.
We next use Chiloe to determine how often islands occur.
We study islands in two systems with 6 \acp{VP} for 3 years
  and 13k \acp{VP} for 3 months.

\textbf{Trinocular:}
We first consider three years of Trinocular data (described in \autoref{sec:data_sources}),
  from 2017-04-01 to 2020-04-01.
We run Chiloe across each VP for this period.

\Cref{tab:island_summary} shows the number of islands per VP
  over this period.
Over the 3 years, all six \acp{VP} see from 1 to 5 islands.
In addition,
  we report as islands some cases even though
  not the \emph{entire} Internet core
  is unreachable.
This apparent discrepancy from our definition
  reflects the limitations of our necessarily
  non-instantaneous measurement of the Internet.
We expect such cases, and perhaps other 12 non-islands where 20\% to 50\% is inaccessible,
  are \emph{short-lived} true islands,
  that are incompletely measured because
  the island recovers before we complete
  an 11~minute-long evaluation of all 5M networks for a full Internet scan
  (see \autoref{sec:island_trinocular_threshold} for details).

\textbf{RIPE Atlas:}
For broader coverage we next consider RIPE Atlas'
  13k \acp{VP} for all of 2021q3~\cite{ripe_ping}.
\oldreviewfix{S22B88}
While Atlas does not scan the whole Internet core,
  they do scan most root DNS servers every 240\,s.
\oldreviewfix{S22B89}
Chiloe would like to observe the whole Internet core, and
  while Trinocular scans 5M /24s,
  it does so with only 6 VPs.
To use RIPE Atlas' VPs,
  we approximate a full scan
  with probes to 12 of the DNS root server systems (G-Root was unavailable in 2021q3).
Although far fewer than 5M networks,
  these targets provide a very sparse sample
  of usually independent destinations since each is independently operated.
Thus we have complementary datasets
  with sparse VPs and dense probing, and
  many VPs but sparse probing.
\oldreviewfix{S22B3}
In other words, to get many VP locations
  we relax our conceptual definition by decreasing our target list.

\begin{figure*}
  \subfloat [Number of islands]{
    \includegraphics[trim=25 0 25 0,clip,width=0.33\linewidth]{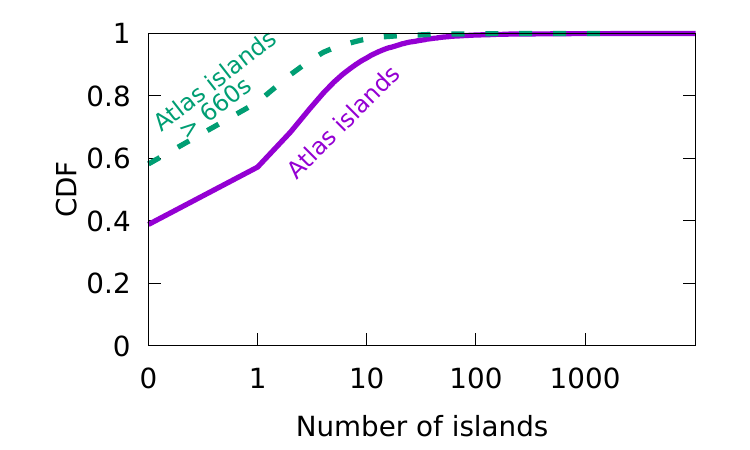}
    \label{fig:islands_per_node}
  }
  \subfloat[Duration of islands]{
    \includegraphics[trim=25 0 25 0,clip,width=0.33\textwidth]{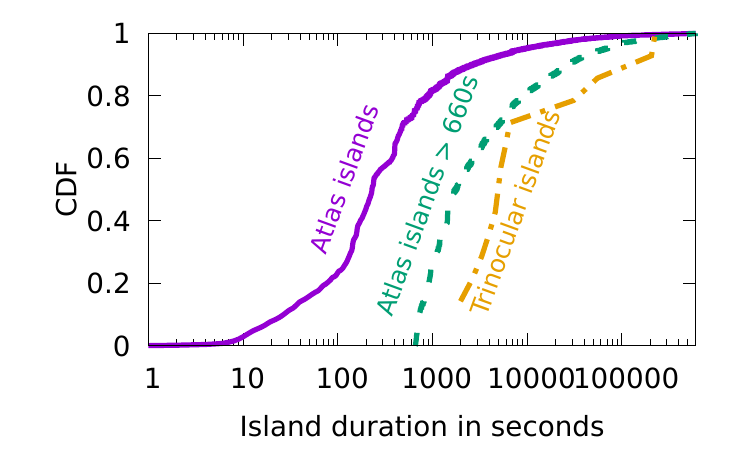}
    \label{fig:island_duration}
  }
  \subfloat[Size of islands]{
    \includegraphics[trim=25 0 25 0,clip,width=0.33\textwidth]{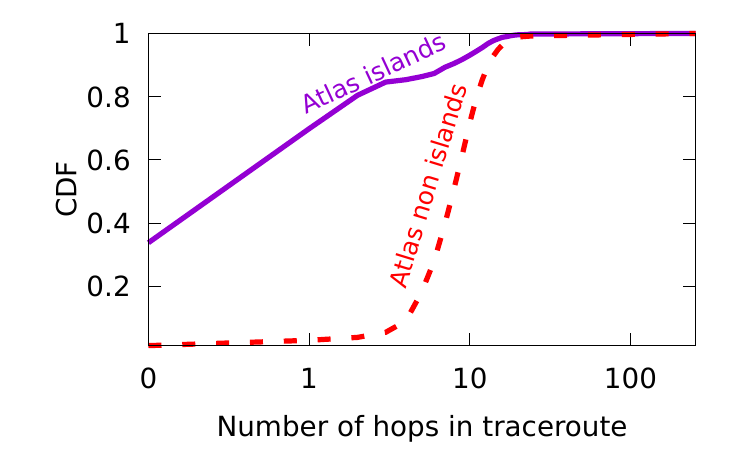}
    \label{fig:island_path_length}
  }
    \captionsetup{type=figure}
    \captionsetup{width=0.9\textwidth}
    \captionof{figure}{CDF of islands detected by Chiloe for data from Trinocular (3 years, Datasets A28-A39) and %
  Atlas (2021q3).}
        \label{fig:a30all_peninsulas_fig7}
\end{figure*}

\autoref{fig:islands_per_node} shows the CDF of the number of islands detected
per RIPE Atlas \ac{VP} during 2021q3.
During this period, 55\% of \acp{VP} observed one or no islands (solid line).
To compare to Trinocular, we consider events longer than 660\,s with the dashed line.
\oldreviewfix{S22B91}
In the figure,
  60\% of \acp{VP} saw no islands, 19\% see one, and the remainder see more.
\oldreviewfix{S22B92}
The annualized island rate of just the most stable \acp{VP} (those that see 2 or less islands)
  is 1.75 islands per year (a lower bound, since we exclude less stable \acp{VP}),
  compared to 1.28 for Trinocular (\Cref{tab:island_summary}).
We see islands are more common in Atlas, perhaps because it includes
  many VPs at home.

We conclude that islands \emph{do} happen,
  but they are rare,
    and at irregular times.
This finding is consistent with importance of the Internet
  at the locations where we run VPs.

\subsection{How Long Do Islands Last?}
\label{sec:islands_duration}

Islands are caused by both brief connectivity loss and
  long-term policy differences,
  so
  we next evaluate island duration.

We compare the distributions of island durations observed from
  RIPE Atlas (the left line) and Trinocular (right) in
  \autoref{fig:island_duration}.
Since Atlas' frequent polling means it detects islands lasting seconds,
  while Trinocular sees only islands of 660\,s or longer,
  we split out Atlas events lasting at least 660\,s
  (middle line).
All measurements follow a similar S-shaped curve,
  but for Trinocular, the curve is truncated at 660\,s.
\oldreviewfix{S22B96}\oldreviewfix{S22B97}
With only 6 VPs, Trinocular sees far fewer events (23 in 3 years compared to 235k in a yearly quarter with Atlas),
  so the Trinocular data is quantized.
In both cases, about 70\% of islands are between 1000 and 6000\,s.
This graph shows that Trinocular's curve is similar in shape to Atlas-660\,s,
  but about $2\times$ longer.
All Trinocular observers are in datacenters,
  while Atlas devices are often at homes,
  so this difference may indicate that datacenter islands are rarer, but harder to resolve.

\subsection{What Sizes Are Islands?}
	\label{sec:islands_sizes}

\oldreviewfix{S22B98}
In \autoref{sec:island} we described different sizes of islands
  starting from as small as an address island,
  as opposed to LAN- or AS-sized islands,
  to country-sized islands potentially capable of partitioning the Internet core.
Here we examine island sizes two ways: first by examining traceroutes,
  and then by considering several large events.

\subsubsection{Island Size via Traceroute}

First we evaluate island sizes
  by counting the number of hops in a traceroute
  sent towards a target outside the island
  before the traceroute fails.

We use traceroutes from RIPE Atlas VPs sent to 12 root DNS servers
  for 2021q3 \cite{ripe_traceroute}.
In \autoref{fig:island_path_length} in green the distribution of the number of hops when traceroute reach their target.
In purple, we plot the distribution of the number of hops of traceroutes that failed to reach the target
for VPs in islands detected in \autoref{sec:how_common_are_islands}.

We find most islands are small, with 70\% at~0 or 1~hop. %
Huge islands (10 or more hops) seen in traceroutes are likely false positives.

\subsubsection{Country-sized Islands}
	\label{sec:country_sized_islands}

We also have some evidence of country-sized islands.

In 2017q3 we observed 8 events when it appears that most or all of China
  stopped responding to external pings.
\autoref{fig:a29all_outagedownup_4096} shows
  the number of /24 blocks that were down over time,
  each spike more than 200k /24s,
  between two to eight hours long.
We found no problem reports on network operator mailing lists,
  so we believe these outages were ICMP-specific and likely did not affect
  web traffic.
Since there were no public reports,
  we assume the millions of computers inside China continued to operate
  and these events were islands and not outages.

We consider these cases examples of China becoming an \emph{ICMP-island}.
We have not seen such large islands since 2017.

\section{Applying These Tools}

\subsection{Outages Given Partial Reachability}
	\label{sec:local_outage_eval}
	\label{sec:representing_the_internet}

We next re-evaluate reports from existing outage detection systems,
  considering how to resolve conflicting information in light of
  our new algorithms.
We compare findings to external information in traceroutes from CAIDA Ark.

\begin{figure*}
\begin{minipage}[b]{.33\linewidth}
  \centering
  \includegraphics[width=6cm]{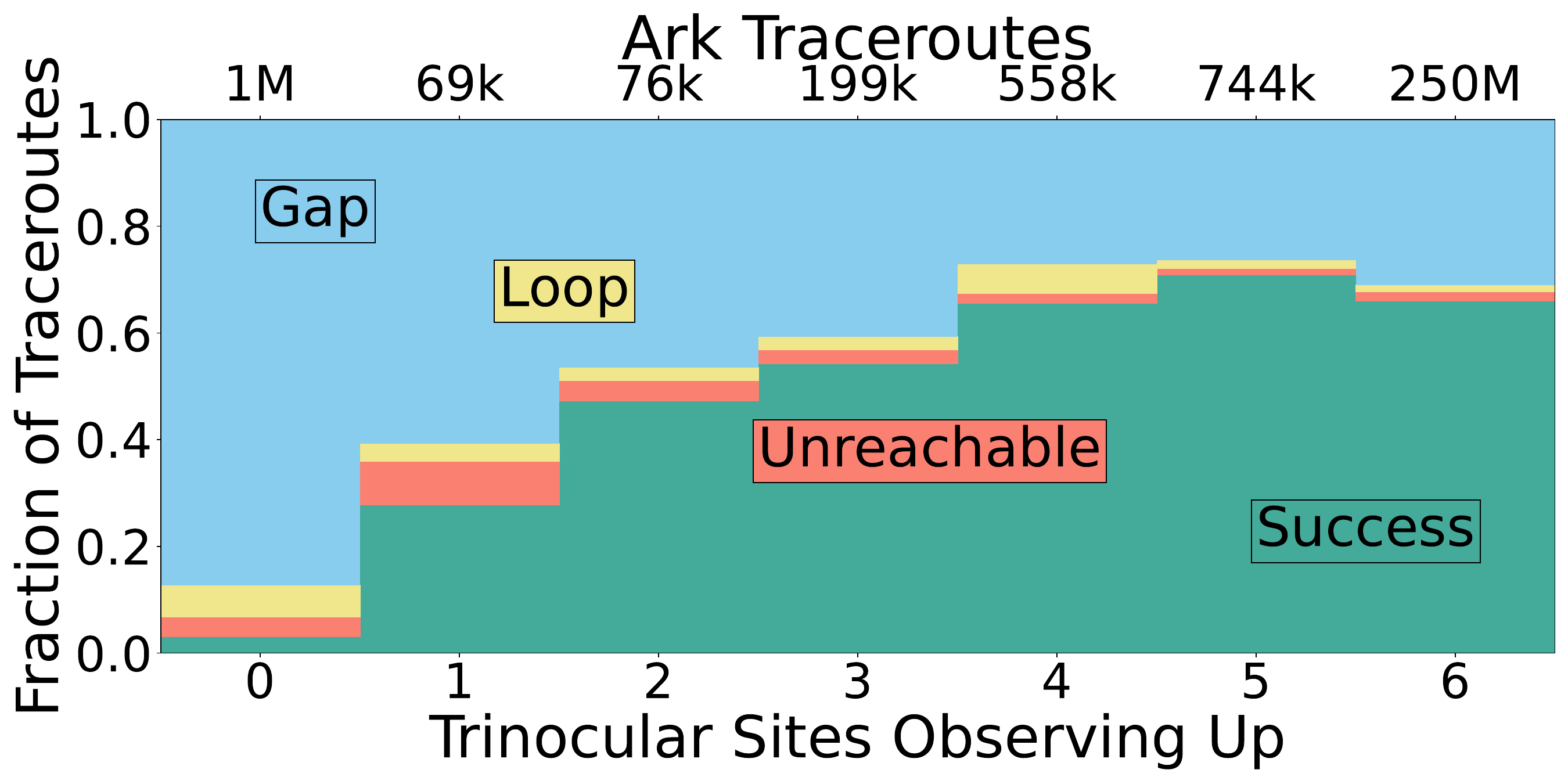}
        \caption{Ark traceroutes sent to targets under partial outages
        (2017-10-10 to -31). Dataset A30.}
  \label{fig:a30all_reach_fraction}
\end{minipage}
\begin{minipage}[b]{.33\linewidth}
    \centering
    \footnotesize
    \resizebox{1\textwidth}{!}{
    \includegraphics[trim=0 0 0 5,clip,width=1\textwidth]{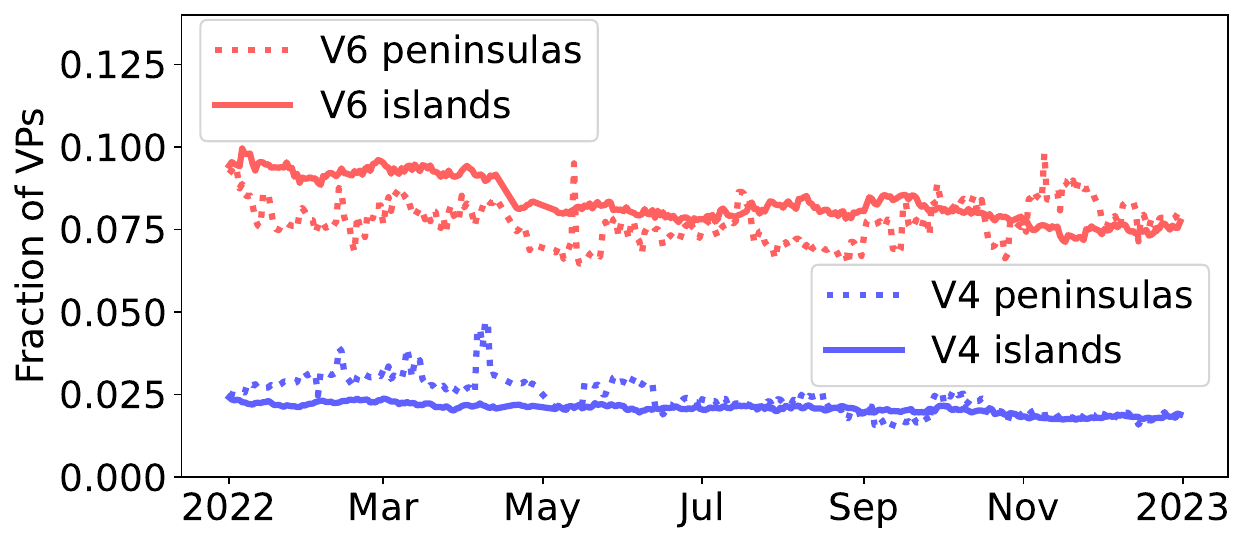}}
        \vspace*{-4mm}
    \captionsetup{type=figure}
    \captionsetup{width=0.9\textwidth}
    \captionof{figure}{Fraction of VPs observing islands and peninsulas for IPv4 and IPv6 during 2022.}
  \label{fig:a49_partial_outages}
\end{minipage}
\begin{minipage}[b]{.33\linewidth}
    \centering
    	\footnotesize
        \resizebox{1\textwidth}{!}{
        \includegraphics[trim=0 0 0 5,clip,width=1\textwidth]{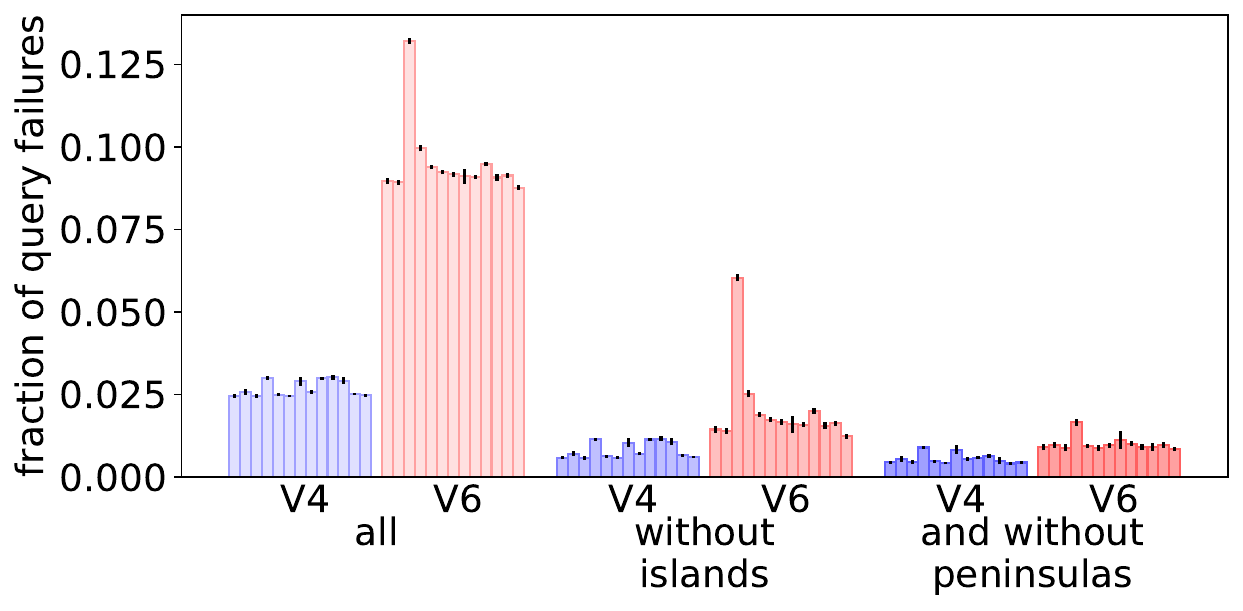}}
        \vspace*{-6mm}
    \captionsetup{type=figure}
    \captionof{figure}{Atlas queries from all available VPs to 13
    Root Servers for IPv4 and IPv6 on 2022-07-23.}
    \label{fig:atlas_revisited}
\end{minipage}
\end{figure*}

\autoref{fig:a30all_reach_fraction} compares Trinocular with
  21 days of Ark topology data, from 2017-10-10 to -31 from all 3 probing teams.
For each Trinocular outage we classify the Ark result as success
  or three types of failure: unreachable, loop, or gap.

Trinocular's 6-site-up case suggests a working network,
  and we consider this case as typical.
However, we see that about 25\% of Ark traceroutes are ``gap'',
  where several hops fail to reply.
We also see about 2\% of traceroutes are unreachable
  (after we discard traceroutes to never reachable addresses).
Ark probes a random address in each block;
  many addresses are non-responsive,
  explaining these.

With 1 to 11 sites up, Trinocular is reporting disagreement.
We see that the number of Ark success cases (the green, lower portion of each bar)
  falls roughly linearly with the number of
  successful observers.
This consistency suggests that Trinocular and Ark are seeing similar behavior,
  and that there is partial reachability---these events
  with only partial Trinocular positive results
  are peninsulas.

Since 5 sites give the same results as all 6,
  single-\ac{VP} failures likely represent problems local to that \ac{VP}.
This data suggests that all-but-one voting will track true outages.

With only partial reachability, with 1 to 4 VPs (of 6),
  we see likely peninsulas.
These cases confirm that partial connectivity is common:
  while there are 1M traceroutes sent to outages where no \ac{VP} can see the target
  (the number of events is shown on the 0 bar),
  there are 1.6M traceroutes sent to partial outages
  (bars 1 to 5),
  and 850k traceroutes sent to definite peninsulas (bars 1 to 4).
This result is consistent with the convergence we see in
\autoref{fig:a30all_peninsulas_duration_oct_nov}.

\subsection{Improving DNSmon Sensitivity}
	\label{sec:dnsmon}

DNSmon~\cite{Amin15a}
  monitors the Root Server System~\cite{RootServers16a}
  from the RIPE Atlas distributed platform~\cite{Ripe15c}.
For years, DNSmon has often reported IPv6 loss rates of 4-10\%.
Since the DNS root is well provisioned and distributed,
  we expect minimal congestion or loss
  and find these values surprisingly high.

RIPE Atlas operators are aware of problems with some Atlas VPs.
Some VPs support IPv6 on their LAN, but not to the global IPv6 Internet---such VPs
  are IPv6 islands.
Atlas periodically tags and culls these VPs
  from DNSmon.
However, our study of DNSmon
  for islands and peninsulas
  improves their results.
Using concepts pioneered here (\autoref{sec:problem} and \autoref{sec:design}),
  we give full analysis in a
    workshop paper~\cite{Saluja22a};
Here we add new data
  showing these results persist for 1 year (\autoref{fig:a49_partial_outages}).

Groups of bars in
  \autoref{fig:atlas_revisited} show query loss
  for each of the 13 root service identifiers,
  as observed
  from all available Atlas VPs (10,082 IPv4, and 5,173 IPv6)
  on 2022-07-23.
(We are similar to DNSmon, but it uses only about 100 well-connected ``anchors'',
  so our analysis is wider.)
The first two groups show loss rates for IPv4 (light blue, left most) and IPv6 (light red),
  showing IPv4 losses around 2\%, and IPv6 from 9 to 13\%.

We apply Chiloe to these VPs, detecting as islands those VPs that
  cannot see \emph{any} of the 13 root identifiers over 24~hours.
(This definition is stricter than regular Chiloe because these VPs attempt only 13 targets,
  and we apply it over a full day to consider only long-term trends.)
The middle two groups of bars show IPv4 and IPv6 loss rates
  after removing VPs that
  are islands.
Without island VPs,
  IPv4 loss rates drop to 0.005 from 0.01, and IPv6 to about 0.01 from 0.06.
These rates represent a more
  meaningful estimate of DNS reliability.
Users of VPs that are IPv6 islands
  will not expect global IPv6,
  and such VPs should not be used for IPv6 in DNSmon.

The third bar in each red cluster of IPv6 is an outlier:
  that root identifier shows 13\% IPv6 loss with all VPs,
  and 6\% loss after islands are removed.
This result is explained by
  persistent routing disputes between Cogent (the operator of C-Root) and Hurricane Electric~\cite{Miller09a}.
Omitting islands (the middle bars) makes this difference much clearer.

Finally we apply Taitao to detect peninsulas.
Peninsulas suggest persistent routing problems;
  they deserve attention from ISPs and root operators.
The darker, rightmost two groups show loss from VPs that are
  neither islands nor peninsulas,
  representing loss if routing problems were addressed.
With this correction C-Root is similar to others,
  confirming that routing disputes account for its different response rates.

This example shows how \emph{understanding partial reachability
  can improve the sensitivity of existing measurement systems}.
Removing islands makes it easy to identify persistent routing problems.
Removing peninsulas makes
  transient changes (perhaps from failure, DDoS, routing)
  more visible.
Each layer of these problems can be interesting,
  but considering each separately,
\oldreviewfix{I23D25}
  the interesting ``signal'' of routing changes
  (appearing in the right two groups in \autoref{fig:atlas_revisited}),
  is hidden under the
  $5\times$ or $9.7\times$ times larger peninsulas and islands (the left two groups).
Improved sensitivity also \emph{shows a need to improve
  IPv6 provisioning},
  since
  IPv6 loss is statistically higher than
  IPv4 loss (compare the right blue and red groups),
  even accounting for known problems.
After sharing the results with root operators and RIPE Atlas,
  two operators adopted them in regular operation.

\section{Related Work}
	\label{sec:related}

Several systems mitigate partial outages.
RON provides alternate-path routing
  around failures for a mesh of sites~\cite{andersen2001resilient}.
Hubble monitors in real-time reachability problems
  when working physical paths exist~\cite{katz2008studying}.
LIFEGUARD, remediates route failures
  by rerouting traffic using BGP to select a
  working path~\cite{katz2012lifeguard}.
While addressing the problem of partial outages,
  these systems do not quantify their duration or scope.

Prior work studied partial reachability, showing
  it is a common transient occurrence
  during routing convergence~\cite{bush2009internet}.
They reproduced partial connectivity with controlled experiments;
  we study it from Internet-wide \acp{VP}.

Internet scanners have examined bias by location~\cite{Heidemann08c},
  more recently looking for policy-based filtering~\cite{wan2020origin}.
  We measure policies with our country specific algorithm, and
  we extend those ideas to defining the Internet core.

Active outage detection systems have encountered partial outages.
Thunderping recognizes a ``hosed'' state with mixed replies,
  but its study is future work~\cite{Schulman11a}.
Trinocular discards partial outages by
  reporting the target block ``up'' if any VP can reach
  it~\cite{quan2013trinocular}.
To the best of our knowledge, prior outage detection systems
  do not consistently report partial outages in the Internet core,
  nor do they study their extent.

Recent groups have studied the policy issues around Internet fragmentation~\cite{Drake16a,Drake22a}, but do not define it.
We hope our definition can fill that need.

\section{Conclusions}

This paper identified partial connectivity as a fundamental challenge in the
  Internet today.
We developed the algorithm Taitao, to find peninsulas of partial connectivity,
  and Chiloe, to find islands.
We showed that partial connectivity events as common as simple outages,
  and use them to to clarify implications of Internet sovereignty
  and to improve outage and DNSmon measurement systems.

\begin{acks}

The authors would like to thank John Wroclawski,
  Wes Hardaker, Ramakrishna Padmanabhan,
  Ramesh Govindan,
  Eddie Kohler,
  Alberto Dainotti,
  and the Internet Architecture Board for their input on
  on an early version of this paper.

The work is
  supported in part by
   the National Science Foundation, CISE
  Directorate, award CNS-2007106 %
  and NSF-2028279. %
The U.S.~Government is authorized to reproduce and distribute
reprints for Governmental purposes notwithstanding any copyright
notation thereon.
\end{acks}

\label{page:last_body}

\bibliographystyle{ACM-Reference-Format}

\appendix

\section{Discussion of Research Ethics}
	\label{sec:research_ethics}

Our work poses no ethical concerns as described in \autoref{sec:introduction}.
We elaborate here.

First, we collect no additional data, but instead
  reanalyze data from several existing sources
  listed in \autoref{sec:data_sources_list}.
Our work therefore poses no additional load on the Internet,
  nor any new risk from data collection.

Our analysis poses no risk to individuals
  because our subject is network topology and connectivity.
There is a slight risk to individuals in that we
  examine responsiveness of individual IP addresses.
With external information, IP addresses can sometimes be traced to individuals,
  particularly when combined with external data sources like DHCP logs.
We avoid this risk in three ways.
First, we do not have DHCP logs for any networks
  (and in fact, most are unavailable outside of specific ISPs).
Second, we commit, as research policy, to not combine
  IP addresses with external data sources
  that might de-anonymize them to individuals.
Finally, except for analysis of specific cases as part of validation,
  all of our analysis is done in bulk over the whole dataset.

We do observe data about organizations such as ISPs,
  and about the geolocation of blocks of IP addresses.
Because we do not map IP addresses to individuals,
  this analysis poses no individual privacy risk.

Finally, we suggest that while our work poses minimal privacy risks
  to individuals,
  to also provides substantial benefit to the community and to individuals.
For reasons given in the introduction
  it is important to improve network reliability and understand
  now networks fail.
Our work contributes to that goal.

Our work was reviewed by the
  Institutional Review Board at our university
  and because it poses no risk to individual privacy,
  it was identified as non-human subjects research
  (USC IRB IIR00001648).

\section{Data Sources and Key Claims }
	\label{sec:data_sources_list}

\begin{table*}
  \resizebox{\textwidth}{!}{%
	\begin{tabular}{llllp{6.5cm}}
	\textbf{Dataset Name} & \textbf{Source} & \textbf{Start Date} & \textbf{Duration} & \textbf{Where Used} \\
	\hline
	\rowcolor[HTML]{DCDCDC}
	internet\_outage\_adaptive\_a28w-20170403 & Trinocular~\cite{trinoculardatasets} & 2017-04-03 & 90 days &  \\
	\rowcolor[HTML]{DCDCDC}
	\quad Polish peninsula subset & & 2017-06-03 & 12 hours & \autoref{sec:island}, \autoref{sec:polish_peninsula_validation} \\
	internet\_outage\_adaptive\_a28c-20170403 & Trinocular & 2017-04-03 & 90 days & \\
	\quad Polish peninsula subset & & 2017-06-03 & 12 hours & \autoref{sec:polish_peninsula_validation}\\
	\rowcolor[HTML]{DCDCDC}
	internet\_outage\_adaptive\_a28j-20170403 & Trinocular & 2017-04-03 & 90 days & \\
	\rowcolor[HTML]{DCDCDC}
	\quad Polish peninsula subset & & 2017-06-03 & 12 hours & \autoref{sec:polish_peninsula_validation}\\
	internet\_outage\_adaptive\_a28g-20170403 & Trinocular & 2017-04-03 & 90 days & \\
	\quad Polish peninsula subset & & 2017-06-03 & 12 hours & \autoref{sec:polish_peninsula_validation}\\
	\rowcolor[HTML]{DCDCDC}
	internet\_outage\_adaptive\_a28e-20170403 & Trinocular & 2017-04-03 & 90 days & \\
	\rowcolor[HTML]{DCDCDC}
	\quad Polish peninsula subset & & 2017-06-03 & 12 hours & \autoref{sec:island}, \autoref{sec:polish_peninsula_validation} \\
	internet\_outage\_adaptive\_a28n-20170403 & Trinocular & 2017-04-03 & 90 days & \\
	\quad Polish peninsula subset & & 2017-06-03 & 12 hours & \autoref{sec:island}, \autoref{sec:polish_peninsula_validation} \\
	\rowcolor[HTML]{DCDCDC}
	internet\_outage\_adaptive\_a28all-20170403 & Trinocular & 2017-04-03 & 89 days & \autoref{sec:chiloe_validation},
  											   \autoref{sec:how_common_are_islands},
    											   \autoref{sec:islands_duration},
  											   \autoref{sec:island_trinocular_threshold} \\
	internet\_outage\_adaptive\_a29all-20170702 & Trinocular & 2017-07-02 & 94 days  & \autoref{sec:island},
											   \autoref{sec:chiloe_validation},
  											   \autoref{sec:how_common_are_islands},
    											   \autoref{sec:islands_duration},
  											   \autoref{sec:island_trinocular_threshold} \\
	\rowcolor[HTML]{DCDCDC}
	internet\_outage\_adaptive\_a30w-20171006 & Trinocular & 2017-10-06 & 85 days & \\
	\rowcolor[HTML]{DCDCDC}
	\quad Site E Island & & 2017-10-23 & 36 hours & \autoref{sec:peninsula_definition}, \autoref{sec:polish_peninsula_validation} \\
	internet\_outage\_adaptive\_a30c-20171006 & Trinocular & 2017-10-06 & 85 days & \\
	\quad Site E Island & & 2017-10-23 & 36 hours & \autoref{sec:polish_peninsula_validation} \\
	\rowcolor[HTML]{DCDCDC}
	internet\_outage\_adaptive\_a30j-20171006 & Trinocular & 2017-10-06 & 85 days & \\
	\rowcolor[HTML]{DCDCDC}
	\quad Site E Island & & 2017-10-23 & 36 hours & \autoref{sec:polish_peninsula_validation} \\
	internet\_outage\_adaptive\_a30g-20171006 & Trinocular & 2017-10-06 & 85 days & \\
	\quad Site E Island & & 2017-10-23 & 36 hours & \autoref{sec:polish_peninsula_validation} \\
	\rowcolor[HTML]{DCDCDC}
	internet\_outage\_adaptive\_a30e-20171006 & Trinocular & 2017-10-06 & 85 days & \\
	\rowcolor[HTML]{DCDCDC}
	\quad Site E Island & & 2017-10-23 & 36 hours & \autoref{sec:peninsula_definition}, \autoref{sec:polish_peninsula_validation} \\
	internet\_outage\_adaptive\_a30n-20171006 & Trinocular & 2017-10-06 & 85 days & \\
	\quad Site E Island & & 2017-10-23 & 36 hours & \autoref{sec:peninsula_definition}, \autoref{sec:polish_peninsula_validation} \\
	\rowcolor[HTML]{DCDCDC}
	internet\_outage\_adaptive\_a30all-20171006 & Trinocular & 2017-10-06 & 85 days  & \autoref{sec:chiloe_validation},
  											   \autoref{sec:how_common_are_islands},
    											   \autoref{sec:islands_duration},
    											   \autoref{sec:site_correlation},
  											   \autoref{sec:island_trinocular_threshold} \\
	\rowcolor[HTML]{DCDCDC}
	\quad   Oct. Nov. subset & & 2017-10-06 & 40 days  &
	\autoref{sec:country_validation},
											   \autoref{sec:peninsula_duration},
											   \autoref{sec:peninsula_size}\\
	\rowcolor[HTML]{DCDCDC}
	\quad	Oct. subset & & 2017-10-10 & 21 days  & \autoref{sec:taitao_validation},
  											   \autoref{sec:representing_the_internet} \\
	internet\_outage\_adaptive\_a31all-20180101 & Trinocular & 2018-01-01 & 90 days  & \autoref{sec:chiloe_validation},
  											   \autoref{sec:how_common_are_islands},
    											   \autoref{sec:islands_duration},
  											   \autoref{sec:island_trinocular_threshold} \\
	\rowcolor[HTML]{DCDCDC}
	internet\_outage\_adaptive\_a32all-20180401 & Trinocular & 2018-04-01 & 90 days  & \autoref{sec:chiloe_validation},
  											   \autoref{sec:how_common_are_islands},
    											   \autoref{sec:islands_duration},
  											   \autoref{sec:island_trinocular_threshold}\\
	internet\_outage\_adaptive\_a33all-20180701 & Trinocular & 2018-07-01 & 90 days  & \autoref{sec:chiloe_validation},
  											   \autoref{sec:how_common_are_islands},
    											   \autoref{sec:islands_duration},
  											   \autoref{sec:island_trinocular_threshold}\\
	\rowcolor[HTML]{DCDCDC}
	internet\_outage\_adaptive\_a34all-20181001 & Trinocular & 2018-10-01 & 90 days  & \autoref{sec:chiloe_validation},
  											   \autoref{sec:how_common_are_islands},
    											   \autoref{sec:islands_duration},
											   \autoref{sec:additional_confirmation},
  											   \autoref{sec:island_trinocular_threshold}\\
	internet\_outage\_adaptive\_a35all-20190101 & Trinocular & 2019-01-01 & 90 days  & \autoref{sec:chiloe_validation},
  											   \autoref{sec:how_common_are_islands},
    											   \autoref{sec:islands_duration},
  											   \autoref{sec:island_trinocular_threshold}\\
	\rowcolor[HTML]{DCDCDC}
	internet\_outage\_adaptive\_a36all-20190401 & Trinocular & 2019-01-01 & 90 days  & \autoref{sec:chiloe_validation},
  											   \autoref{sec:how_common_are_islands},
    											   \autoref{sec:islands_duration},
  											   \autoref{sec:island_trinocular_threshold}\\
	internet\_outage\_adaptive\_a37all-20190701 & Trinocular & 2019-01-01 & 90 days  & \autoref{sec:chiloe_validation},
  											   \autoref{sec:how_common_are_islands},
    											   \autoref{sec:islands_duration},
  											   \autoref{sec:island_trinocular_threshold}\\
	\rowcolor[HTML]{DCDCDC}
	internet\_outage\_adaptive\_a38all-20191001 & Trinocular & 2019-01-01 & 90 days  & \autoref{sec:chiloe_validation},
  											   \autoref{sec:how_common_are_islands},
    											   \autoref{sec:islands_duration},
  											   \autoref{sec:island_trinocular_threshold}\\
	internet\_outage\_adaptive\_a39all-20200101 & Trinocular & 2020-01-01 & 90 days  & \autoref{sec:chiloe_validation},
  											   \autoref{sec:how_common_are_islands},
    											   \autoref{sec:islands_duration},
  											   \autoref{sec:island_trinocular_threshold}\\
	\rowcolor[HTML]{DCDCDC}
	internet\_outage\_adaptive\_a41all-20200701 & Trinocular & 2020-07-01 & 90 days  & \autoref{sec:peninsula_locations} \\
	\hline
	prefix-probing & Ark~\cite{CAIDA07b} \\
	\quad Oct. 2017 subset & & 2017-10-10 & 21 days & \autoref{sec:taitao_validation}, \autoref{sec:representing_the_internet} \\
	\quad 2020q3 subset & & 2020-07-01 & 90 days & \autoref{sec:peninsula_locations}\\
	\rowcolor[HTML]{DCDCDC}
	probe-data     & Ark & & & \\
	\rowcolor[HTML]{DCDCDC}
	\quad Oct 2017 subset     & & 2017-10-10 & 21 days & \autoref{sec:taitao_validation}, \autoref{sec:representing_the_internet} \\
	\rowcolor[HTML]{DCDCDC}
	\quad 2020q3 subset     & & 2020-07-01 & 90 days & \autoref{sec:peninsula_locations}\\
	\hline
	routeviews.org/bgpdata & Routeviews~\cite{routeviews} & 2017-10-06 & 40 days
    &
      \autoref{sec:country_validation},
  								      \autoref{sec:polish_peninsula_validation} \\
	\hline
	\rowcolor[HTML]{DCDCDC}
	Atlas Recurring Root Pings (id: 1001 to 1016) & Atlas~\cite{ripe_ping} & 2021-07-01 & 90 days & \autoref{sec:peninsula_frequency},
													\autoref{sec:islands_duration} \\

	\end{tabular}
  }
   \caption{All datasets used in this paper.}
   \label{tab:datasets}
\end{table*}

\autoref{tab:datasets} provides a full list of datasets used in this paper
  and where they may be obtained.

\begin{table*}
\begin{tabular}{lll}
 & \textbf{claim} & \textbf{support} \\
\multirow{3}{*}{\rotatebox[origin=c]{90}{\footnotesize examples}}
 & IPv4 islands exist (\autoref{sec:island}) & example Trinocular/2017q2 \\
 & IPv6 peninsulas exist (\autoref{sec:peninsula_definition}) & public news~\cite{ipv6peeringdisputes,google_cogent,cloudflare_he}, DNSmon~\cite{dnsmon}, looking glass~\cite{he_looking_glass,cogent_looking_glass} \\
 & IPv4 peninuslas exist (\autoref{sec:peninsula_definition}) & example Trinocular/2017q4, Ark~\cite{CAIDA07b}, traceroutes and routeviews (\autoref{sec:polish_peninsula_validation}) \\
  \hline
\multirow{3}{*}{\rotatebox[origin=c]{90}{\footnotesize validation}}
 & Taitao correctness (\autoref{sec:taitao_validation}) & Trinocular/2017q4, validated with Ark \\
 & Chiloe correctness (\autoref{sec:chiloe_validation}) & Trinocular/2017q1 to 2020q1 \\
 & 6 Trinocular sites are independent (\autoref{sec:site_correlation}) & Trinocular/2017q4 \\
  \hline
\multirow{5}{*}{\rotatebox[origin=c]{90}{\small observations}}
 & Peninsulas are common (\autoref{sec:peninsula_frequency}) & Trinocular/2017q1, 2018q4, 2020q3 \\
 & Some peninsulas are long-lived (\autoref{sec:peninsula_duration}) & Trinocular/2017q3 to 2020q1 and RIPE Atlas (2021q3) \\
 & Islands are common (\autoref{sec:how_common_are_islands}) &  Trinocular/2017q3 to 2020q1 and RIPE Atlas (2021q3) \\
 & Most island are hours or less (\autoref{sec:islands_duration}) &  Trinocular/2017q3 to 2020q1 and RIPE Atlas (2021q3) \\
 & Most islands are small (\autoref{sec:islands_sizes}) & RIPE Atlas/2021q3 \\
   \hline
\multirow{4}{*}{\rotatebox[origin=c]{90}{\small applications}}
 & peninsulas can clarify outages (\autoref{sec:local_outage_eval})) & Trioncular/2017q4 with Ark (21 days) \\
 & DNSmon sensitivity can improve (\autoref{sec:dnsmon}) & DNSmon/2022 (uses RIPE Atlas) \\
\end{tabular}
\caption{Key observations made in the paper and which datasets support each.}
  \label{tab:claims}
\end{table*}

\autoref{tab:claims} summarizes the key
  observations this paper makes about the Internet,
  and what datasets support each.
The paper body provides all of our key claims
  and validates them with multiple data sources
  and broad, Internet-wide data as observed
  from 6 (Trinocular), about 150 (Ark),  and about 10k (RIPE Atlas)
  locations.
We emphasize that \emph{all} our key results
  use data from multiple data sources.
Some graphs emphasize data source from Trinocular,
  since it provides very broad coverage,
  all key validation and observational results are supported
  with data from either Ark or RIPE Atlas.
All of our trends are verified with
  Trinocular data from 6 sites scanning millions of networks,
  and confirmed by data from many sites scanning less frequently
    (Ark, with 150 sites scanning millions of networks daily)
  or less completely (RIPE Atlas, with 10k sites scanning 13 destinations).

Our core quantitative results use Trinocular data,
  because it is the only currently available data source
  that provides very broad coverage (5M IPv4 blocks)
  with sufficient frequency (updates every 11 minutes)
  to approach an Internet wide view.
However,
  we emphasize that we have validated
  these conclusions as described in \autoref{sec:validation}
  against other Ark data.
These results strongly confirm that
  Taitao and Chiloe true positives and true negatives are correct,
  and suggest there are relatively few true positives.
{2024-06-06}
Trinocular's observations from 6 VPs
  many underrepresent peninsulas
  (Taitao likely has an unknown number of false negatives),
Finally, in many cases, we validate observations with multiple quarters
  of Trinocular data to show that the results are consistent over time.

The utility of core results are also strongly confirmed
  by applying Taitao and Chiloe
  to DNSmon in \autoref{sec:dnsmon}.
DNSmon provides observations from about 10k VPs,
  although they contact 13 sites (the DNS root servers)
  and therefore require a relaxed version of the algorithms.
These results show peninsulas and islands have profound effects
  on DNSmon reports about root-server-system reliability,
  confirming the Trinocular Internet-wide data.

\section{Additional Analysis of Potential Missed Peninsulas}
	\label{sec:false_negative_details}

In \autoref{sec:taitao_validation}
  we suggest that it is difficult to use Ark
  to predict a peninsula because its methodology
  causes it to usually probe a non-responsive address.
This result follows from Ark's design as a topology discovery system,
  not partition detection---as a result,
  the majority of Ark probes fail to reach a target,
  even for a block that is reachable.
(We do not mean this statement as a negative comment about Ark.
It is well suited for topology discovery and building router-level
  network topologies.
This problem occurs when we do our best to reuse Ark data do
  validate network partitions.)

\textbf{Comparing Ark and Trinocular data:}
We reuse Ark to validate network partitions for two reasons:
  first, each traceroute targets one address (prefix probing, the .1, and team probing a random address),
  not multiple addresses.
Second, Ark visits each block on average every 36 minutes,
  compared to Trinocular's observation every 2 minutes (6 observers every 11 minutes).

The biggest challenge is that Ark probes either only one destination address.
By contrast, Trinocular probes up to 15 addresses, stopping on success,
  and those addresses are drawn from addresses that have previously responded.
Predicted addresses respond 49\% of the time,
  while a random address responds in less than 1\% of blocks (\cite{Fan10a}, Figure 7).
Probing 15 addresses is successful more than 90\% of time
  for 5/6ths of responsive blocks (\cite{quan2013trinocular}, Figure 6),
  while probing .1 is responsive only in 1/6th of responsive blocks,
  and other addresses respond even less frequently.
As a result, we expect 4/6ths of Ark targets to be non-responsive,
  even when the block is reachable.
While this choice does not limit Ark for building router-level topologies,
  prediction of peninsulas from multiple Ark attempts
  with mixed results, \emph{almost always reflects Ark targeting
  a non-responsive address, not an actual peninsula}.

Ark's less frequent probing is a second factor.
In aggregate, Ark probes a target block every 36 minutes (40 teams,
  each trying each block once per day),
  while Trinocular probes 18x more frequently (6 observers,
  each probing every block every 11 minutes).
As a result, combinations of Ark observations are often from different times,
  blurring outages and peninsulas.

A ``positive'' is Taitao detecting a peninsula that Ark confirms,
  and a negative is Taitao not detecting a peninsula
  (showing all positive or all negative)
  that Ark has mixed results about.
However, since 90\% of Ark results are negative because
  it chooses a target address that doesn't respond,
  \emph{most} Ark results are mixed and are false indications of peninsulas.
For this reason, we consider Ark reports about false negatives untrustworthy,
  and we discount Ark's evaluation of metrics that depend on the false negative rate
    (such as recall, but not precision).
In other words, Ark can confirm what Taitao finds,
  but it is insufficient to evaluate what Taitao misses.

\textbf{Impliciations in comparing Ark and Taitao:}
We observe that it is quite challenging to compare Taitao and Ark.
We would like to take Ark as ``ground truth''
  from more VPs (150, not just 6),
  but it is designed for topology discovery, not peninsula detection,
  makes it an imperfect tool for this application.

However, we suggest it does provide useful validation
  for true positives, true negatives, and false positives,
  as we describe next.

\textbf{Implications of potential false negatives:}
While we cannot trust Ark's judgment of false negatives,
  it is likely that \emph{some} are correct---Ark likely sees
  some partitions that Taitao misses when using only Trinocular as a data source.
We expect that there are many very small peninsulas (micro-peninsulas),
  and that adding
  more VPs will increase the number of these micro-peninsulas.
As a thought experiment, \emph{every} computer that
  can route to a LAN using public IP addresses,
  but lacks a global route, is a peninsula.

\begin{figure}
	\begin{center}
            \includegraphics[trim=25 0 25 0,clip,width=0.99\columnwidth]{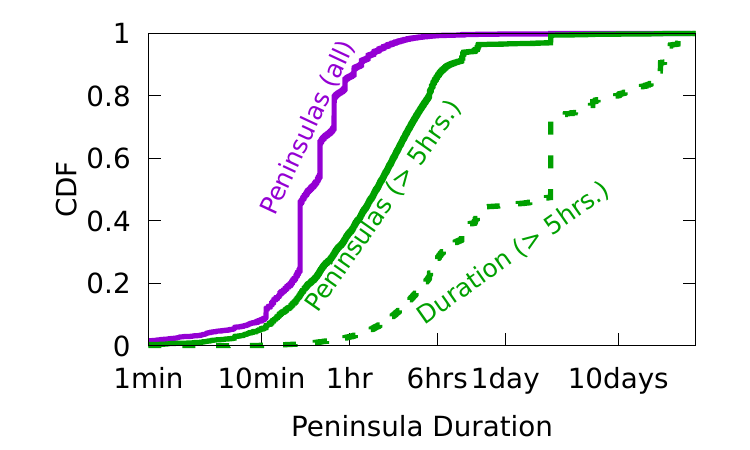}
	\end{center}
	\caption{Cumulative peninsulas and peninsula duration. Dataset A30, 2017q4.}
          \label{fig:a30_partial_outages_duration_cdf}
\end{figure}

We align our claims with this potential:
  we claim there are many peninsulas, at least more than there are outages,
  as supported by \autoref{fig:a30all_peninsulas_duration_oct_nov}.
Although we think there are not many large peninsulas,
  we recognize there are many,
  and some are long-lasting (\autoref{fig:a30_partial_outages_duration_cdf}).
If some or many of the false negatives suggested by Ark
  are actual peninsulas,
  implies that understanding peninsulas is even more important
  than we predict.

\textbf{Implications of true positives, true negatives, and false positives:}
While care must be taken when using Ark as ``ground truth''
  to judge false negatives (peninsulas missed by Taitao),
  it is much more promising to test other conditions.

True positives are
  peninsulas identified by Taitao because it sees conflicting VPs.
For these cases,
  presence of conflicting Ark data is consistent with Taitao.
Even if Ark has excessive last-hop failures,
  its large number of observers suggest that some may be correctly unreachable,
  and the presence of reachability from some Ark VPs confirms
  partial reachability.

True negatives last a long time, so in these cases
  all Taitao VPs reach the target,
  and the long duration of ``all Taitao VPs reach''
  makes it very likely some Ark VP reaches the target.

Finally, the very small number when Taitao reports all down
but Ark shows reachability (6 cases) confirms that in a few cases,
  6 sites are not enough to see all peninsulas.

\section{Validation of the Polish Peninsula}
	\label{sec:polish_peninsula_validation}

We define peninsulas in \autoref{sec:peninsula_definition}
  and present a example peninsula we discovered
  through our algorithms.
That example illusrates the concept,
  but
  here we expand on that example to providing additional
  data from BGP that support our interpretation of the event.

On 2017-10-23, for a period of 3 hours starting at 22:02Z,
  five Polish \acp{AS}
  had 1716 blocks that were unreachable from five \acp{VP}
  while the same blocks remained reachable from a sixth \ac{VP}.

\begin{figure}
  \includegraphics[width=1\columnwidth]{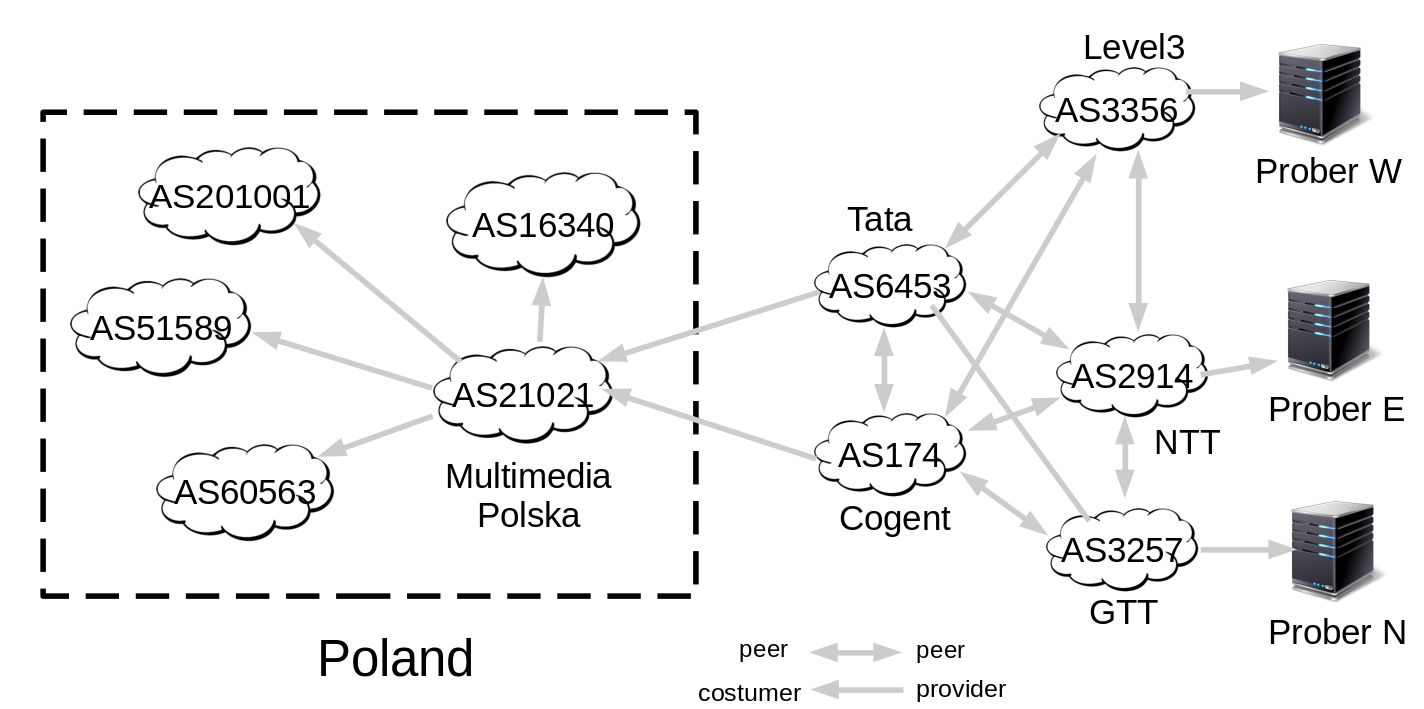}
  \caption{AS level topology during the Polish peninsula.}
  \label{fig:polish_peninsula}
\end{figure}

\oldreviewfix{S21B2}
\autoref{fig:polish_peninsula} shows the AS-level relationships
  at the time of the peninsula.
Multimedia Polska (AS21021, or \emph{MP}) provides service to the other 4 ISPs.
MP has two Tier-1 providers:
  Cogent (AS174) and Tata (AS6453).
Before the peninsula, our VPs see MP
  through Cogent.

\begin{figure}
  \includegraphics[width=0.9\columnwidth]{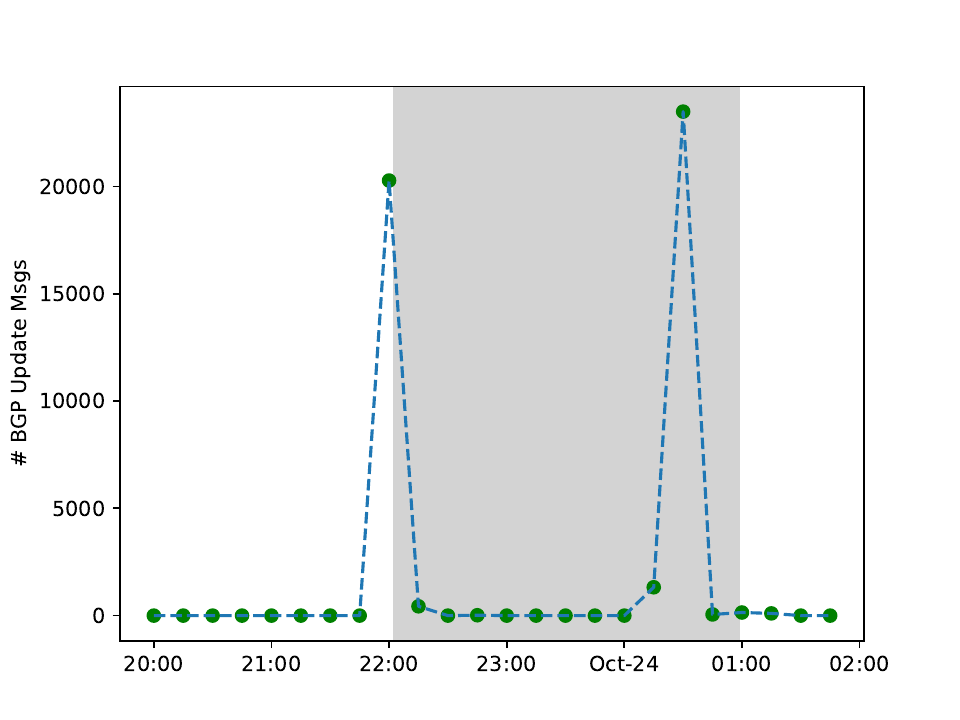}
	\caption{ BGP update messages sent for affected Polish blocks starting
    2017-10-23t20:00Z. Data source: RouteViews. }
  \label{fig:rv_time_plot}
\end{figure}

At event start, we observe many BGP updates (20,275) announcing
and withdrawing routes to the affected blocks(see~\autoref{fig:rv_time_plot}).
These updates correspond to Tata announcing MP's prefixes.
Perhaps MP changed its peering
  to prefer Tata over Cogent,
  or the MP-Cogent link failed.

Initially, traffic from most VPs continued through Cogent
  and was lost; it did not shift to Tata.
One \ac{VP} (W) could reach MP
  through  Tata for the entire event,
  proving MP was connected.
After 3 hours, we see another burst of BGP updates (23,487 this time),
  making MP reachable again from all VPs.

\begin{figure}
  \includegraphics[width=1\columnwidth]{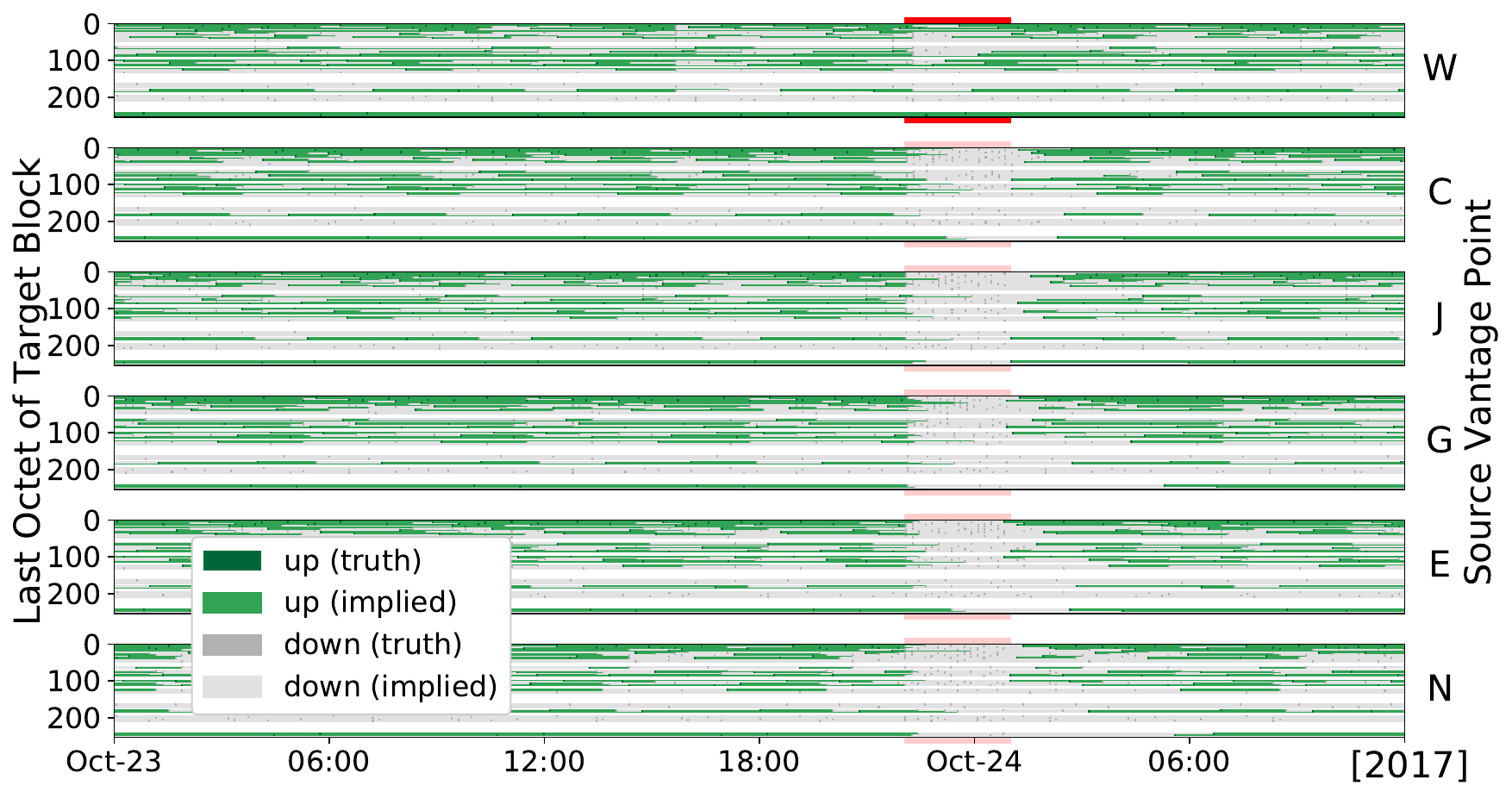}
  \caption{A block
            (80.245.176.0/24)
      showing a 3-hour peninsula accessible only from \ac{VP} W (top bar)
      and not from the other five \acp{VP}.  Dataset: A30.}
  \label{fig:a30all_raw_50f5b000_6sites}
\end{figure}

In \autoref{fig:a30all_raw_50f5b000_6sites} we provide data from our 6 external VPs,
where W is uniquely capable of reaching the target block, thus living in the
same peninsula.

\begin{table*}
    \begin{tabular}{c  c  c  >{\raggedright\arraybackslash}m{120mm}}
            \toprule
src block  &    dst block   &   time        &   traces \\
\hline
c85eb700   &    50f5b000    &   1508630032  & q, 148.245.170.161, 189.209.17.197, 189.209.17.197, 38.104.245.9, 154.24.19.41, 154.54.47.33, 154.54.28.69, 154.54.7.157, 154.54.40.105, 154.54.40.61, 154.54.43.17, 154.54.44.161, 154.54.77.245, 154.54.38.206, 154.54.60.254, 154.54.59.38, 149.6.71.162, 89.228.6.33, 89.228.2.32, 176.221.98.194
\\
\hline
c85eb700   &     50f5b000   &     1508802877 &
q, 148.245.170.161, 200.38.245.45, 148.240.221.29 \\
\bottomrule
    \end{tabular}
    \caption{Traces from the same Ark VPs (mty-mx) to the same destination
            before and during the event block}
    \label{tab:traceroutes}
\end{table*}

We further verify this event by looking at traceroutes.
During the event we see 94 unique Ark VPs attempted 345 traceroutes to the affected blocks.
Of the 94 VPs, 21 VPs (22\%) have their last responsive
  traceroute hop in the same \ac{AS} as the
  target address, and 68 probes (73\%) stopped before reaching that \ac{AS}.
\autoref{tab:traceroutes} shows traceroute data from a single CAIDA Ark VP
  before and during the peninsula described in \autoref{sec:peninsula_definition} and
  \autoref{fig:a30all_50f5b000_accum}. %
This data confirms the block was reachable from some locations and not others.
During the event, this trace breaks at the last hop within the source \ac{AS}.

\section{Additional Details about Islands}

We define islands and give examples in \autoref{sec:island}.
Here we supplement those results with  examples of country-sided islands (\autoref{sec:country_sized_islands}),
  and details about a specific island around one of our \acp{VP} (\autoref{sec:additional_island}).
We also show the raw data we use to justify our choice of 50\% unreachability
  to define islands in Trinocular (\autoref{sec:island_trinocular_threshold}).

\subsection{Validation of the Sample Island}
\label{sec:additional_island}

In \autoref{sec:island} we reported an island affecting a /24 block where
\ac{VP} E lives.
During the time of the event, E was able to successfully probe addresses within
the same block, however, unable to reach external addresses.
This event started at 2017-06-03t23:06Z, and
can be observed in ~\autoref{fig:islands_plot_down_fraction}.

\begin{figure}
  \includegraphics[width=1\columnwidth]{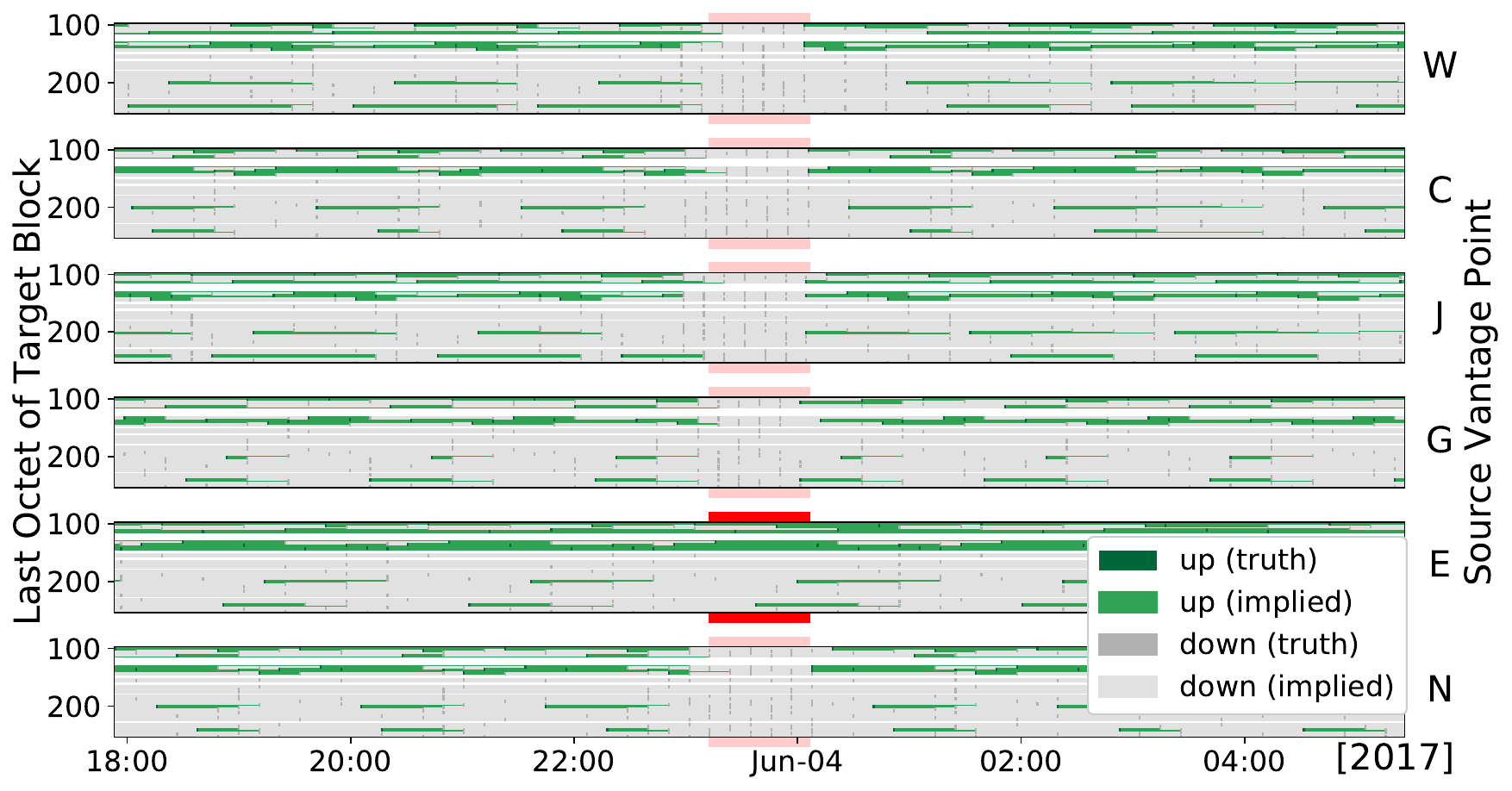}
  \caption{A block
    showing a 1-hour island for this block and \ac{VP} E, while other five
    VPs cannot reach it.}
  \label{fig:a28all_raw_417bca00_6sites}
\end{figure}

Furthermore, no other \ac{VP} was able to reach the affected block for the
time of the island as shown in \autoref{fig:a28all_raw_417bca00_6sites}.

\subsection{Longitudinal View Of Islands}
\label{sec:island_trinocular_threshold}

\begin{figure*}
        \begin{center}
          \includegraphics[width=0.85\linewidth]{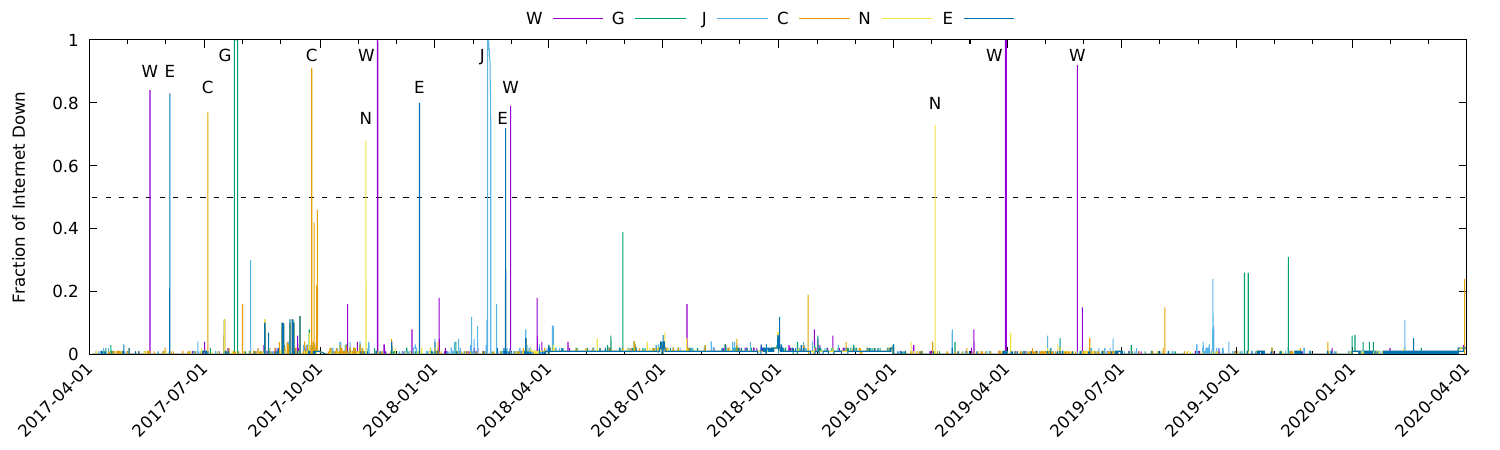}
        \end{center}
        \caption{Islands detected across 3 years using six \acp{VP}.
        Datasets A28-A39.}
  \label{fig:islands_plot_down_fraction}
\end{figure*}

We first consider three years of Trinocular data (described in \autoref{sec:data_sources}),
  from 2017-04-01 to 2020-04-01.
\autoref{fig:islands_plot_down_fraction}
  shows the fraction of the Internet that is
  reachable
  as a dotted line at the 50\% threshold that Chiloe uses to detect an island (\autoref{sec:chiloe}).
We run Chiloe across each VP for this period.

\section{Stability of Results over Time}
\label{sec:2020}

Our paper body uses Trinocular measurements for 2017q4 because
  this time period had six active VPs,
  allowing us to make strong statements about how multiple perspectives help.
Those three months of data provide evidence of result stability,
  but those observations are now several years old.
Because IP address allocation and partial reachability
  are associated with organizational policy (of ICANN and the \acp{RIR})
  and business practices of thousands of ISPs,
  we expect them to change relatively slowly.
Here we verify this assumption,
  showing our results from 2017 hold in 2018 and 2020.
They do---we find quantitatively similar results between 2017
  for number and sizes of peninsulas in 2018q4 in \autoref{sec:additional_confirmation},
  and duration in 2020q3 in \autoref{sec:additional_peninsula_duration},
  confirming these results in \autoref{sec:evaluation} hold.

\subsection{Additional Confirmation of the Number of Peninsulas}
\label{sec:additional_confirmation}

\begin{figure*}
\adjustbox{valign=b}{\begin{minipage}[b]{.33\linewidth}
    \includegraphics[width=1\linewidth]{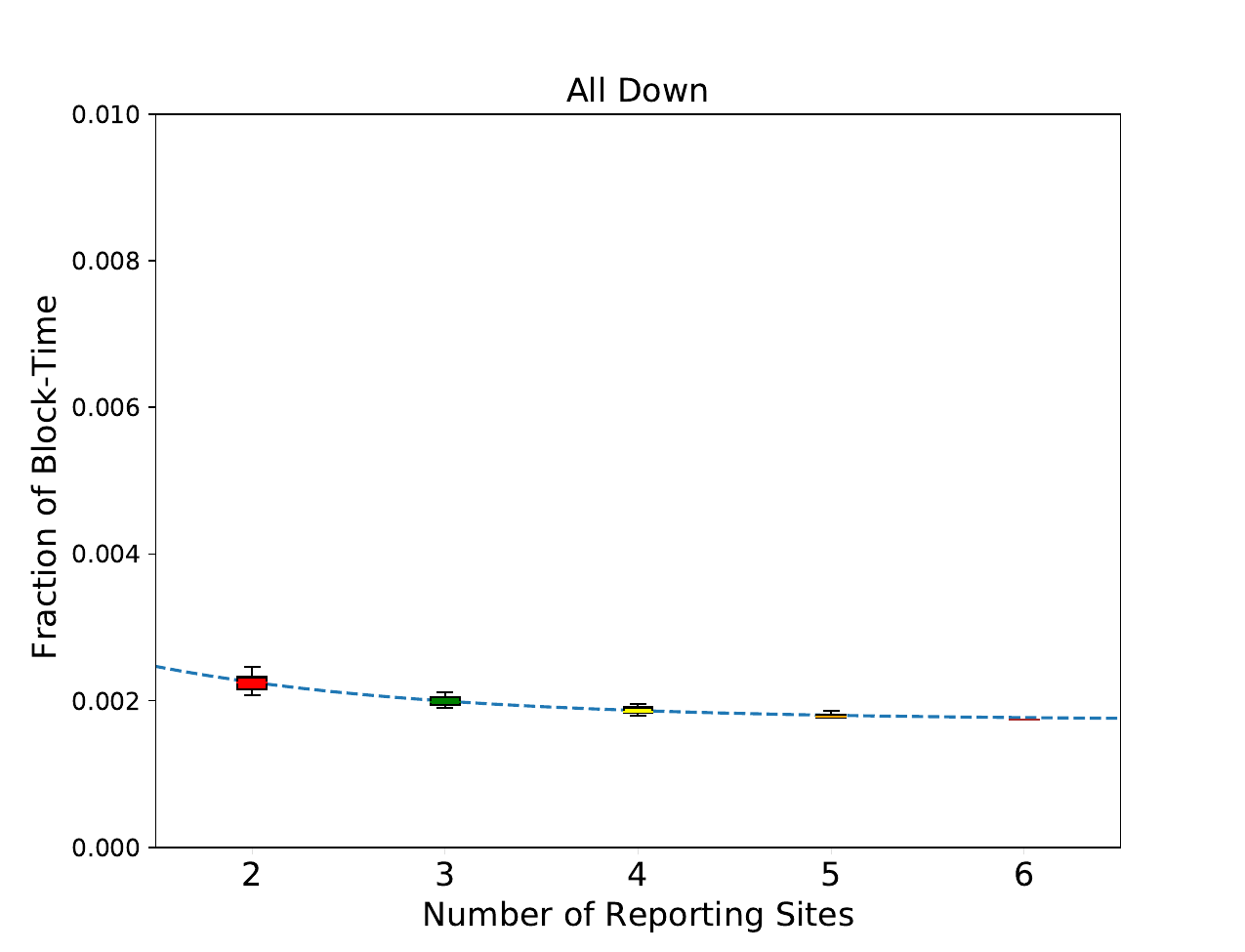}
\end{minipage}}
\adjustbox{valign=b}{\begin{minipage}[b]{.33\linewidth}
    \includegraphics[width=1\linewidth]{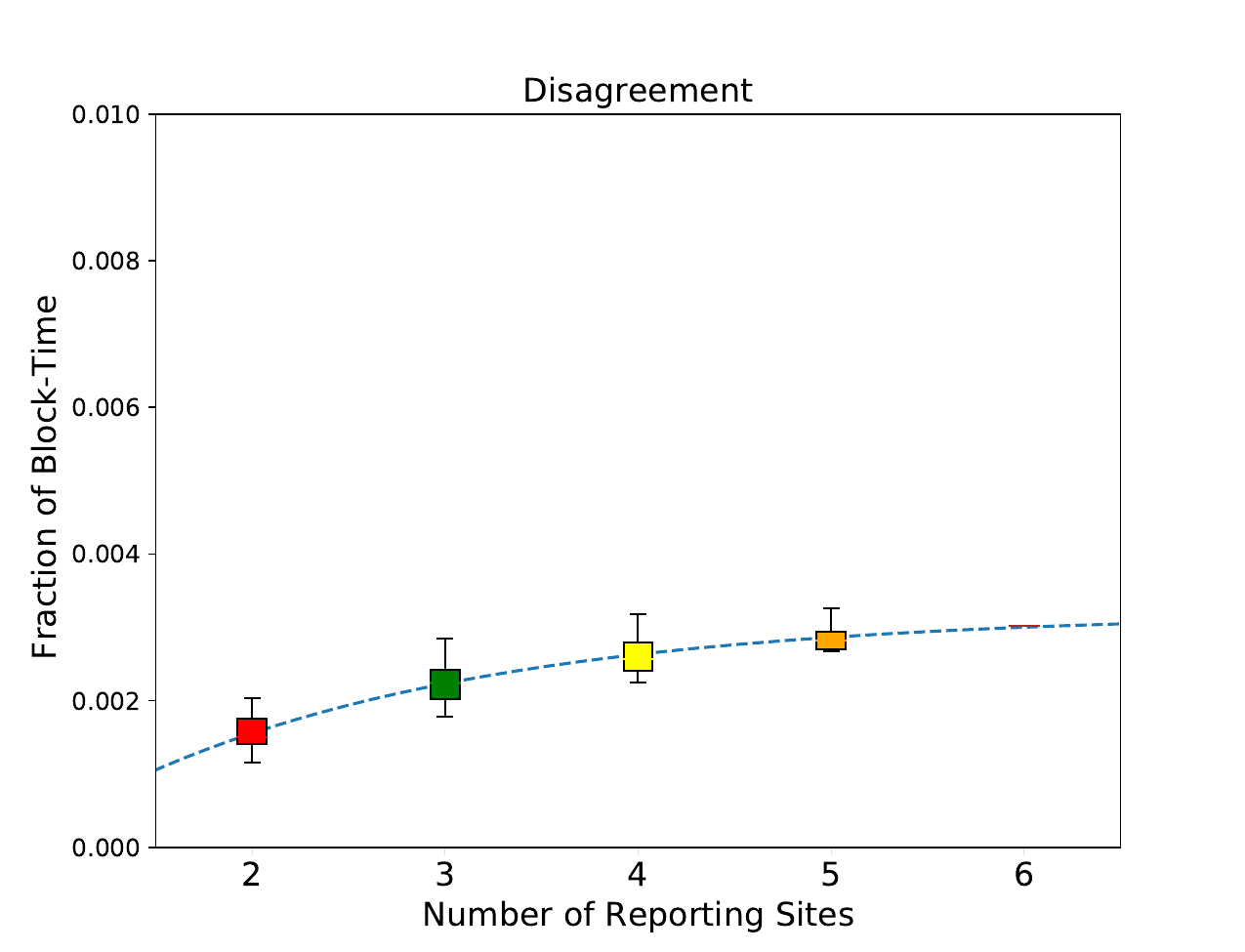}
\end{minipage}}
\adjustbox{valign=b}{\begin{minipage}[b]{.33\linewidth}
    \includegraphics[width=1\linewidth]{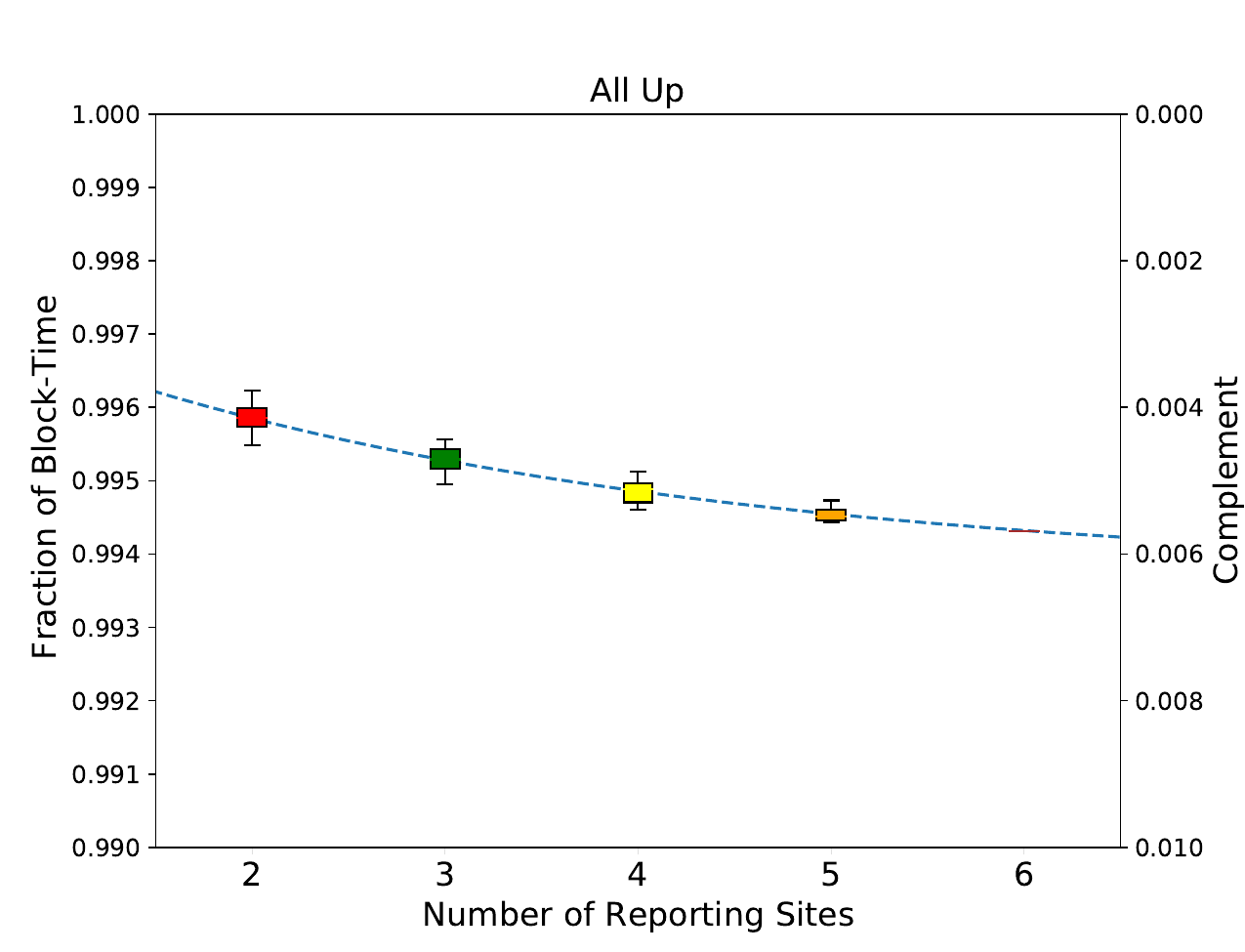}
\end{minipage}}
\caption{Distribution of block-time fraction over sites reporting all down
(left), disagreement (center), and all up (right), for events longer than five
hour. Dataset A34, 2018q4.}
\label{fig:a34all_peninsulas_duration_box}
\end{figure*}

In \autoref{sec:peninsula_frequency}
  we quantify how common peninsulas are.
Here we confirm we see qualitatively similar results,
  but in Trinocular from 2018q4 data.

In \autoref{fig:a34all_peninsulas_duration_box} we confirm,
  that with more \acp{VP} more peninsulas are discovered,
  providing a better view of the Internet's overall state.

\emph{Outages (left) converge after 3 sites},
  as shown by the fitted curve and decreasing variance.
Peninsulas and all-up converge more slowly.

At six \acp{VP}, here we find and even higher difference between all down and
disagreements.
Confirming that peninsulas are a more pervasive problem than outages.

\subsection{Additional Confirmation of Peninsula Duration}
	\label{sec:additional_peninsula_duration}

In \autoref{sec:peninsula_duration} we characterize peninsula duration for
2017q4,
  to determine peninsula root causes.
To confirm our results, we repeat the analysis, but with 2020q3 data.

\begin{figure*}
\begin{center}
  \subfloat [Cumulative events (solid) and duration (dashed)]{
    \includegraphics[width=.63\columnwidth]{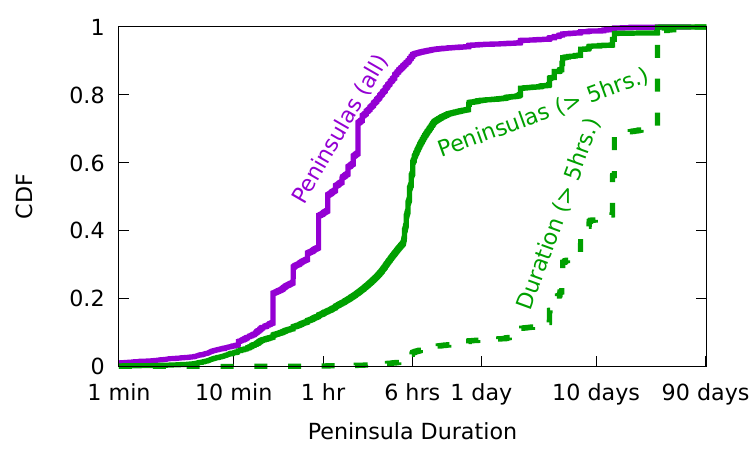}
    \label{fig:a41_partial_outages_duration_cdf}
  }
\quad
  \subfloat[Number of Peninsulas]{
    \includegraphics[width=0.65\columnwidth]{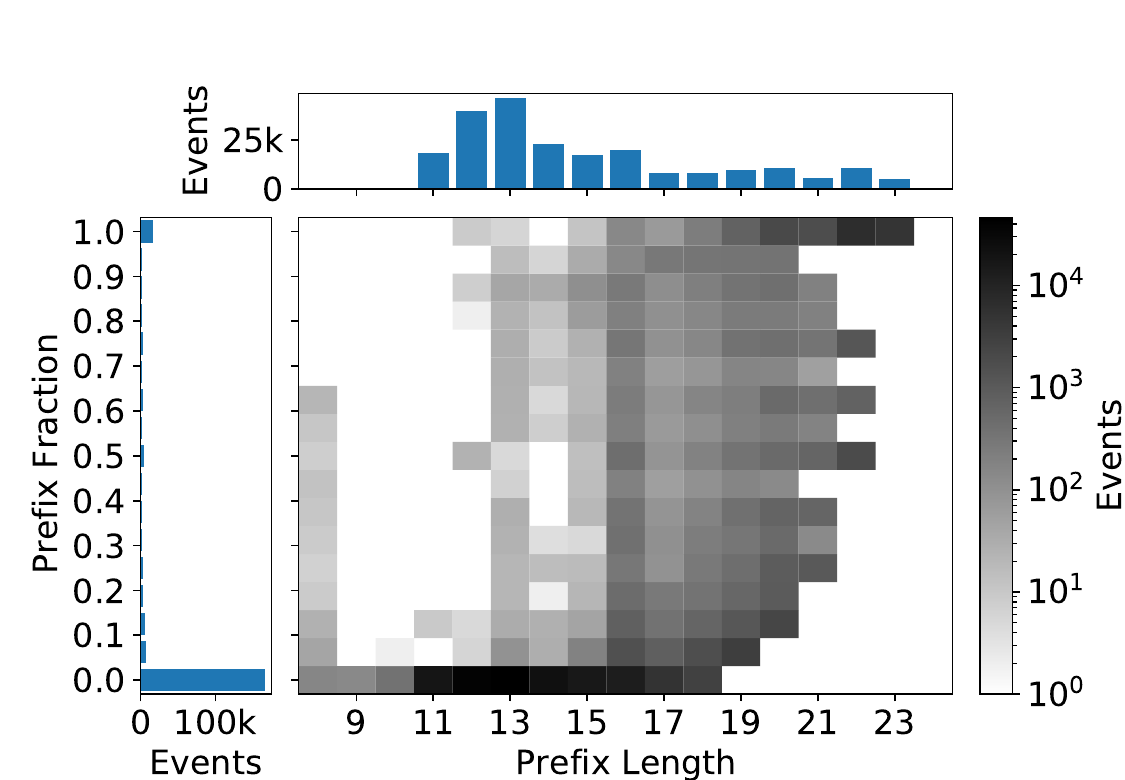}
    \label{fig:a41all_blocks_in_prefix_prefix_fraction_heatmap}
  }
\quad
  \subfloat[Duration fraction]{
    \includegraphics[width=0.65\columnwidth]{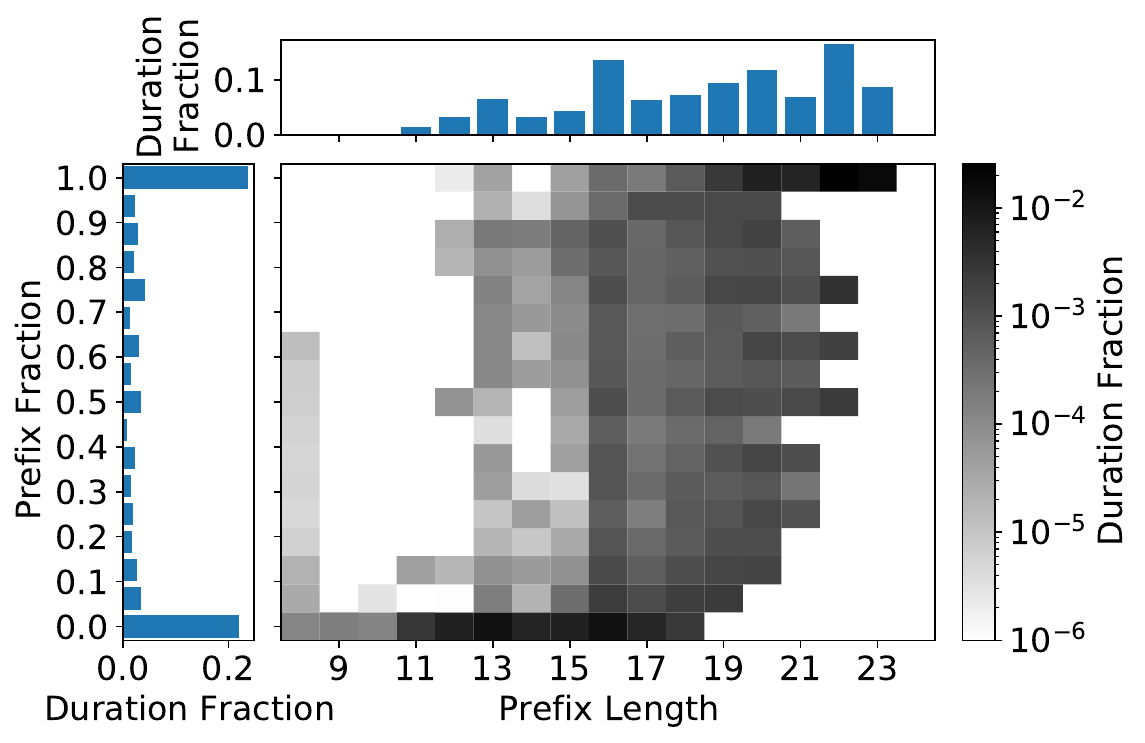}
    \label{fig:a41all_blocks_in_prefix_prefix_fraction_heatmap_duration}
  }
\end{center}
\caption{Peninsulas measured with per-site down events longer than 5 hours during 2020q3. Dataset A41.}
          \label{fig:sim_connection_rollup}
\end{figure*}

As \autoref{fig:a41_partial_outages_duration_cdf} shows,
similarly, as in our 2017q4 results,
  we see that there are many very brief peninsulas (from 20 to 60 minutes).
These results suggest that while the Internet is robust,
there are many small connectivity glitches.

Events shorter than two rounds (22 minutes),
  may represent BGP transients or failures due to random packet loss.

The number of multi-day peninsulas is small,
However, these represent about 90\% of all peninsula-time.
Events lasting a day are long-enough that can be debugged by human network operators,
  and events lasting longer than a week are long-enough that
    they may represent policy disputes.
Together, these long-lived events suggest that
  there is benefit to identifying non-transient peninsulas
  and addressing the underlying routing problem.

\subsection{Additional Confirmation of Size}

In \autoref{sec:peninsula_size} we discussed the size of peninsulas measured as
a fraction of the affected routable prefix.
In the latter section, we use 2017q4 data.
Here we use 2020q3 to confirm our results.

\autoref{fig:a41all_blocks_in_prefix_prefix_fraction_heatmap} shows the peninsulas
per prefix fraction, and \autoref{fig:a41all_blocks_in_prefix_prefix_fraction_heatmap_duration}.
Similarly,
  we find that while small prefix fraction peninsulas are more in numbers,
  most of the peninsula time is spent in peninsulas covering the whole prefix.
This result is consistent with long lived peninsulas being caused by policy
choices.

\newpage \hspace*{1ex}
\label{page:last_page}

\end{document}